\documentclass[10pt,conference]{IEEEtran}
\usepackage{cite}
\usepackage{amsmath,amssymb,amsfonts}
\usepackage{algorithmic}
\usepackage{graphicx}
\usepackage{textcomp}
\usepackage{xcolor}
\usepackage[hyphens]{url}
\usepackage{fancyhdr}
\usepackage{hyperref}
\usepackage{tikz}

\usepackage[table]{xcolor}
\usepackage{booktabs}
\usepackage{makecell}
\usepackage{array}

\newcommand*\circled[1]{\tikz[baseline=(char.base)]{
            \node[shape=circle,draw,inner sep=1.2pt] (char) {\small{#1}};}}
            
\definecolor{dodgeblue}{RGB}{30,144,255}  
\definecolor{pink}{RGB}{255,102,153}      
\definecolor{moreorange}{RGB}{255,153,51} 
\definecolor{green}{RGB}{33, 171, 128}    
\definecolor{pinkred}{RGB}{255,60,60}     
\definecolor{purple}{RGB}{204,102,255}    

\newcommand{\mem}[0]{HBM-CO}

\newcommand{\hpcayear}{2026}
 
\newcommand{\hpcasubmissionnumber}{1268}

\title{RPU -- A Reasoning Processing Unit} 

\def\hpcacameraready{} 

\newcommand\hpcaauthors{Matthew Joseph Adiletta, Gu-Yeon Wei and David Brooks}
\newcommand\hpcaaffiliation{Harvard University}

\author{
  \ifdefined\hpcacameraready
    \IEEEauthorblockN{\hpcaauthors{}}
      \IEEEauthorblockA{
        \hpcaaffiliation{}
      }
  \else
    \IEEEauthorblockN{\normalsize{HPCA \hpcayear{} Submission
      \textbf{\#\hpcasubmissionnumber{}}} \\
      \IEEEauthorblockA{
        Confidential Draft \\
        Do NOT Distribute!!
      }
    }
  \fi 
}

\fancypagestyle{camerareadyfirstpage}{%
  \fancyhead{}
  
  \fancyhead[C]{
    \ifdefined\aeopen
    \parbox[][12mm][t]{13.5cm}{\hpcayear{} IEEE International Symposium on High-Performance Computer Architecture (HPCA)}    
    \else
      \ifdefined\aereviewed
      \parbox[][12mm][t]{13.5cm}{\hpcayear{} IEEE International Symposium on High-Performance Computer Architecture (HPCA)}
      \else
      \ifdefined\aereproduced
      \parbox[][12mm][t]{13.5cm}{\hpcayear{} IEEE International Symposium on High-Performance Computer Architecture (HPCA)}
      \else
      \parbox[][0mm][t]{13.5cm}{}
    \fi 
    \fi 
    \fi 
    \ifdefined\aeopen 
      \includegraphics[width=12mm,height=12mm]{ae-badges/open-research-objects.pdf}
    \fi 
    \ifdefined\aereviewed
      \includegraphics[width=12mm,height=12mm]{ae-badges/research-objects-reviewed.pdf}
    \fi 
    \ifdefined\aereproduced
      \includegraphics[width=12mm,height=12mm]{ae-badges/results-reproduced.pdf}
    \fi
  }
  \fancyfoot[C]{}
}
\begin{document}
\maketitle

\ifdefined\hpcacameraready 
  \thispagestyle{camerareadyfirstpage}
  \pagestyle{empty}
\else
  \thispagestyle{plain}
  \pagestyle{plain}
\fi

\newcommand{\hpcaheight}{0mm}
\ifdefined\eaopen
\renewcommand{\hpcaheight}{12mm}
\fi


\begin{abstract}

Large language model (LLM) inference performance is increasingly bottlenecked by the memory wall.
While GPUs continue to scale raw compute throughput, they struggle to deliver scalable performance for memory bandwidth bound workloads.
This challenge is amplified by emerging reasoning LLM applications, where long output sequences, low arithmetic intensity, and tight latency constraints demand significantly higher memory bandwidth.
As a result, system utilization drops and energy per inference rises, highlighting the need for an optimized system architecture for scalable memory bandwidth.

To address these challenges we present the Reasoning Processing Unit (RPU), a chiplet-based architecture designed to address the challenges of the modern memory wall. 
RPU introduces: 
(1) A Capacity-Optimized High-Bandwidth Memory (HBM-CO) that trades capacity for lower energy and cost; 
(2) a scalable chiplet architecture featuring a bandwidth-first power and area provisioning design; and 
(3) a decoupled microarchitecture that separates memory, compute, and communication pipelines to sustain high bandwidth utilization. 
Simulation results show that RPU performs up to 45.3× lower latency and 18.6× higher throughput over an H100 system at ISO-TDP on Llama3-405B.

\end{abstract}

\section{Introduction}

Low-latency inference is critical for reasoning LLMs, which may generate thousands of tokens before reaching a final answer~\cite{wei_chain--thought_2023, ballon_relationship_2025, chen_not_2025, wang_thoughts_2025}.
Without fast token generation, reasoning tasks become slow, limiting the adoption of this powerful inference approach.
Therefore, we introduce the Reasoning Processing Unit (RPU), a new chiplet-based system architecture that delivers orders of magnitude higher memory bandwidth than  today’s compute-centric designs, addressing the core latency bottleneck in reasoning LLMs.

The first step toward low-latency LLM inference is to separate prefill and decode onto different systems, as in Dynamo~\cite{elmeleegy_introducing_2025} and the Splitwise execution model~\cite{patel_splitwise_2024}. 
Prefill is compute-bound and highly parallel, making it efficient to run on today’s GPU architectures. 
Decode, in contrast, is inherently sequential and latency-sensitive.
When prioritizing latency, decode systems must operate at low batch sizes, often as low as one. 
This low-batch regime is not a choice but a necessity, for two key reasons.

First, each query must compute attention sequentially during decode, so larger batch sizes lead to higher latencies. 
Long sequences are common in reasoning models, which makes this effect even more pronounced.
Second, latency-optimized techniques like speculative decoding~\cite{leviathan_fast_2023} are only effective at low batch sizes~\cite{gandhi_distil-whisper_2023}.
As a result, small batch sizes are essential for low latency.
However, low-batch inference exacerbates the memory wall problem.
For example, Figure 1 shows low-batch decode operates far below the H100’s compute roofline.

\begin{figure}
    \centering
    \includegraphics[width=\linewidth]{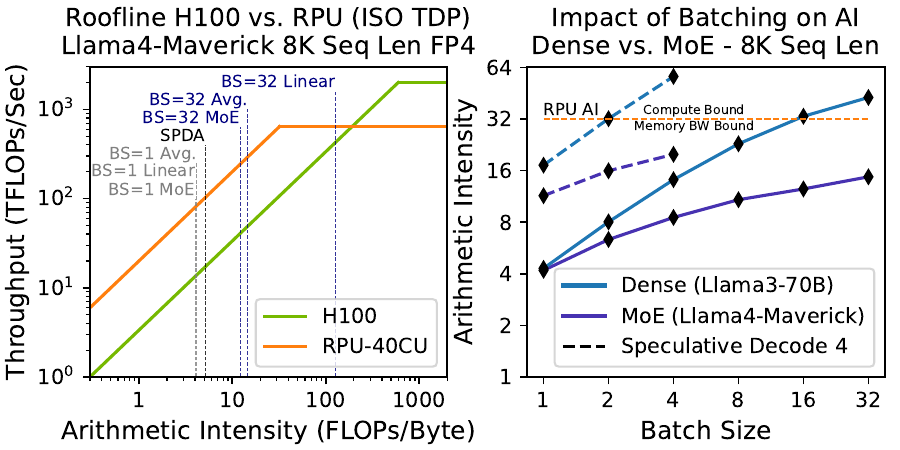}
    \caption{RPU provides higher memory bandwidth than H100, which is required for low-latency decoding. Even up to BS=32, arithmetic intensity remains low, but requires the RPU to execute kernels which straddle the roofline.}
    \label{fig:roofline}
\end{figure}




System architects have responded to the decode bottleneck by cramming more HBM stacks into each package as the simplest way to chase bandwidth~\cite{choquette_nvidia_2023, smith_amd_2024, tirumala_nvidia_2024, kaplan_intel_2024}. 
However, HBM was not designed for inference. 
Its popularity was driven primarily by GPU training, pushing a roadmap focused on high bandwidth coupled with high capacity to store massive training datasets and large model checkpoints to feed dense compute architectures. 
Now, as applications shift toward inference, HBM’s high capacity per module drives up energy and cost, undermining scalability for low-latency inference.

\textit{Challenge 1 -- Energy and Cost of Memory.}
In streaming workloads, over 74\% of memory device energy is spent moving data across long internal wires and TSVs within the DRAM stack~\cite{oconnor_fine-grained_2017, moon_advanced_2023}. 
Higher capacity HBMs increase internal paths, driving up energy per bit. 
This creates a fundamental mismatch for low-reuse 
streaming workloads, where minimizing energy per bit is critical for efficiency.
Similarly, the cost of next-generation AI accelerators is increasingly dominated by HBM memory~\cite{patel_ai_2023, patel_memory_2024}. 
Cost scales with capacity, as larger stacks require more silicon area and complex packaging.

The capacity required per socket depends on how a model is deployed (e.g., a single node, across a rack~\cite{noauthor_nvidia_2025}, or at datacenter scale~\cite{jouppi_tpu_2023}).
If a model fits in one GPU, doubling the number of sockets doubles bandwidth, but it also halves the capacity needed per socket.
We call this the \textit{memory overprovisioning paradox}: a system-level design inefficiency where scaling out for bandwidth to achieve lower latency overprovisions capacity and drives up system cost and energy.
The correct architectural approach is to provision memory based on the scale of the intended deployment.
But with HBM as the only high-bandwidth memory available~\cite{kim_present_2024}, architects are locked into a costly compromise -- \textit{buying bandwidth with capacity}.


\textbf{\textit{Contribution 1 -- \mem: An HBM-style memory design which can be optimized to the bandwidth-capacity needs of low-batch LLMs, addressing energy and cost challenges of the memory wall (Section~\ref{sec:bw-per-cap}).}} 
\mem{} retains HBM’s internal bandwidth architecture and shoreline bandwidth, but selectively reduces capacity-driving structures such as ranks, banks and subarrays to optimize for the capacity needs of low-latency inference. 
These changes require minimal modification to the HBM stack, making \mem{} a practical and manufacturable design point. 
This yields higher bandwidth per dollar and up to 2.4× energy efficiency than conventional HBM, despite a higher cost per GB.
We quantify these tradeoffs using an analytical modeling approach based on~\cite{oconnor_fine-grained_2017}, which estimates energy per bit and cost per module from wire-length scaling trends from HBM core-die floorplans~\cite{lee_134_2024, ryu_16_2023, park_192-gb_2023}.

Modularizing memory with \mem{} unlocks a new design regime.
Instead of provisioning large, monolithic capacity per socket, systems can scale bandwidth and capacity across many smaller stacks. 
But this flexibility introduces a new architectural challenge: how do we build a compute fabric that can glue together these \mem{} modules, deliver the desired bandwidth, and stay within tight power and packaging limits? 
The answer requires rethinking how power and area are provisioned in modern accelerators.

\textit{Challenge 2 -- Power and Area Provisioning for A Scalable Compute Fabric:}
Today's HBM-based accelerators (e.g., H100~\cite{choquette_nvidia_2023}, MI300x~\cite{smith_amd_2024}) allocate only 30-40\% of system thermal design power (TDP) to memory interfaces, meaning that memory-bound workloads leave a large fraction of available power underutilized.
These systems are also reticle-limited, which favor dense compute, but this leads to overprovisioned arithmetic and cache resources during bandwidth bound workloads.
Additionally, memory bandwidth scales with die perimeter, not area, because each HBM stack requires a dense ring of high-speed IOs along the chip edge (shoreline)~\cite{canakci_good_2025}. 
Reticle-limited designs minimize this perimeter-to-area ratio, which directly conflicts with the need for scalable bandwidth.

Alternatively, accelerators like Cerebras~\cite{wang_cerebras_2024} and Groq~\cite{noauthor_groq_2024} avoid shoreline constraints by using SRAM as main memory. 
However, SRAM’s low density drives up power, cost, and infrastructure overheads. 
For example, Cerebras requires four WSEs to host a 70B model, and Groq needs hundreds of processors.
In contrast, DRAM provides the density to support large models at far lower system footprint.

\textbf{\textit{Contribution 2 -- RPU: A modular chiplet-based architecture that dedicates more power to memory interfaces and optimizes the compute-to-bandwidth ratio for low-latency token generation (Section~\ref{sec:comp-fabric}).}}
The RPU embraces emerging trends in package-level integration to achieve scalable memory bandwidth. 
Rather than concentrating compute in a monolithic die, the RPU distributes compute across many smaller chiplets. 
As a result, for the same compute die area, the RPU exposes nearly 10× more memory IO shoreline than the H100 (600mm vs. 60mm), enabling tighter coupling between memory and logic.
Four compute chiplets are co-packaged with eight \mem{} stacks, following package-level integration strategies pioneered by modern GPU architectures~\cite{choquette_nvidia_2023, smith_amd_2024, prabhakar_sambanova_2024, jouppi_tpu_2023, gomes_ponte_2022}. 
Packages are composed at the board level, increasing bandwidth until blade-level power envelopes.

In parallel, the RPU reprovisions power and area. 
The RPU dedicates 70-80\% of power to memory interfaces. 
This keeps the memory bandwidth bound power near the peak power and enables over 2× bandwidth at ISO TDP versus compute-centric GPUs. 
The RPU also aligns the compute-to-bandwidth ratio. 
By removing underutilized compute and cache resources, it improves area efficiency and reduces die cost.
As shown in Figure~\ref{fig:roofline}, these RPU design choices shift the roofline down and to the left, which is a better fit for low-latency inference. 


The RPU system-architecture make more bandwidth available; however, roofline availability is not the same as utilization. 
Realizing the full potential of the RPU system requires rethinking how we sustain bandwidth at scale.

\textit{Challenge 3 -- Utilizing the Newly Available Bandwidth:}
Today's systems struggle to use memory bandwidth effectively, particularly in low-batch LLM token generation~\cite{zhang_llmcompass_2024, kundu_performance_2024}.
This is because small, distributed weight matrices limit streaming bandwidth~\cite{liu_performance_2025, prabhakar_sambanova_2024}.
For example, the fused \textit{gate/up} projection MLP layer in Llama4-Maverick contains just 168 million parameters \texttt{(5k×32k)}. 
When \textit{column-sharded}~\cite{wang_overlap_2022} across devices with TB/s of bandwidth, the roofline model predicts  runtimes in the tens of microseconds. 
In practice, kernel launch and tensor-parallel communication latencies are often of similar magnitude, which limit bandwidth utilization~\cite{liu_performance_2025}.

\textbf{\textit{Contribution 3 -- Reasoning Core: A decoupled pipeline microarchitecture and custom ISA that fully saturates available memory bandwidth (Section~\ref{sec:micro-architecture}).}}
Achieving high bandwidth utilization requires careful orchestration of data movement. 
The RPU Reasoning Core microarchitecture addresses this by separating dataflow for memory, compute, and network into independent pipelines, connected by on-chip buffers and coordinated through programmable pipeline arbiters. 
This decoupling allows each pipeline to make forward progress based on data readiness, without stalling on global barriers. 
At batch size 1 \texttt{(BS=1)}, the RPU saturates memory bandwidth and achieves roofline performance. 
On the other hand, batch size 32 \texttt{(BS=32)}, contains kernels that straddle the roofline, as shown in Figure 1, alternating between memory-bound SDPA and MoE layers and compute-bound Linear layers. 
Decoupled pipelines enable the RPU to absorb this phase imbalance into the buffer hierarchy and sustain throughput at the workloads average arithmetic intensity (AI). 

\textbf{\textit{Contribution 4: An end-to-end simulation framework for LLM inference on RPU (Sections~\ref{sec:simulation}-~\ref{sec:strong-scaling}).}}
We developed a simulation framework, combining RTL-modeled compute kernels, an analytical energy model of \mem{} memory, and an event-driven simulator~\cite{sanchez_zsim_2013, akram_survey_2019} that executes compiled LLM workloads via a custom RPU ISA.
This framework captures transient dataflows, pipeline utilization, and synchronization stalls, which enables detailed architectural comparisons against modern GPUs.
Compared to an H100 at ISO-TDP, RPU achieves up to 45.3× lower latency and 18.6× higher throughput on Llama3-405B at similar system cost.

\begin{figure*}
    \centering
    \includegraphics[width=.69\linewidth]{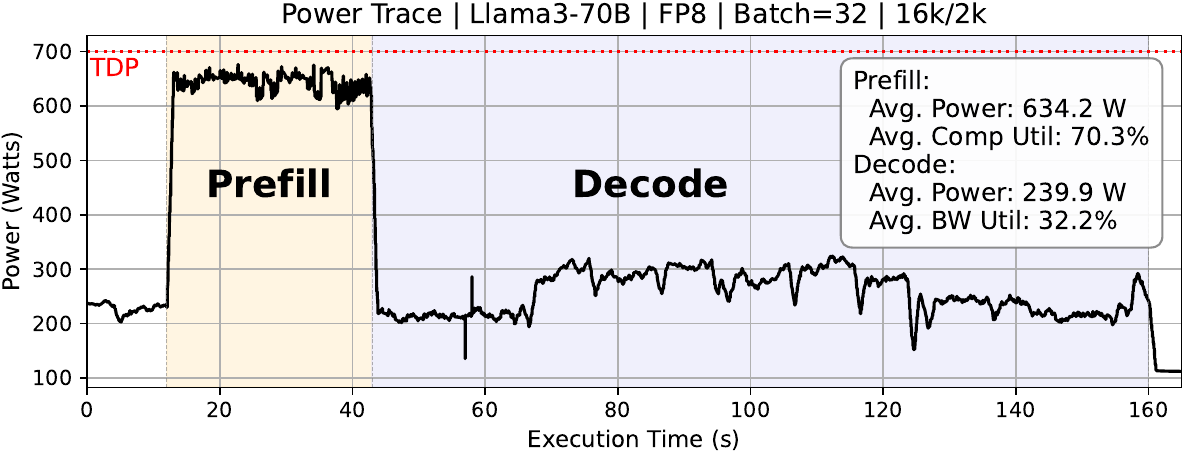}
    \includegraphics[width=.3\linewidth]{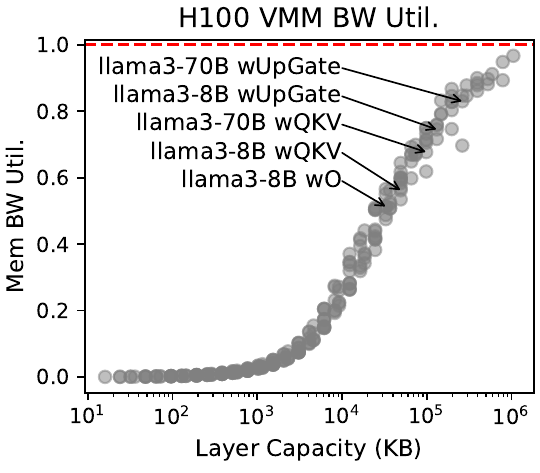}
    \caption{Power and utilization characterization of H100 using NVML. 
    Left: Power trace during distributed inference (4xH100) of Llama3-70B (Batch=32). 
    Right: Isolated kernel profiling for memory bandwidth utilization across batch sizes and matrix dimensions (BF16). }
    \label{fig:h100-characterization}
\end{figure*}

\begin{figure}
    \centering
    \includegraphics[width=\linewidth]{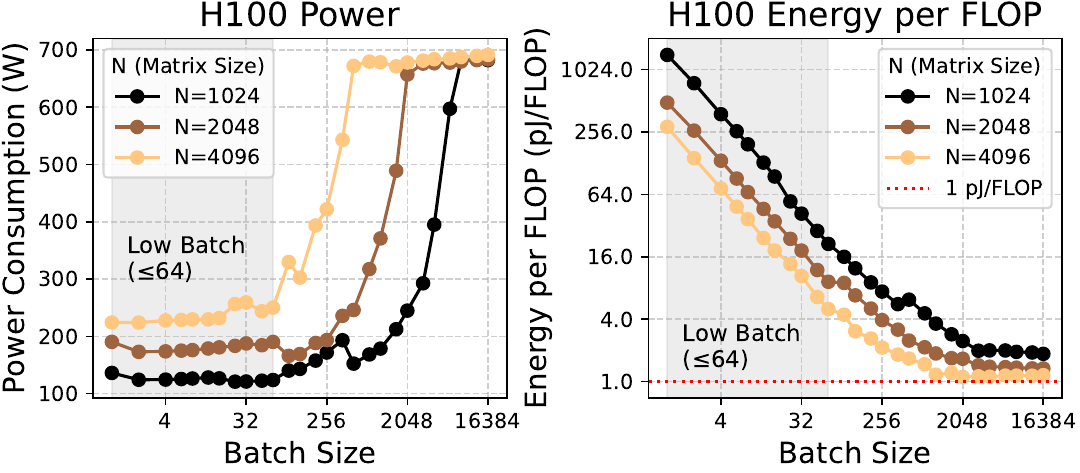}    \caption{Isolated kernel profiling for power consumption and energy efficiency across batch sizes and matrix dimensions (BF16).}
    \label{fig:isolated-kernel-profiling}
\end{figure}

\section{Motivation for a Decode-Optimized System}\label{sec:motivation}

To illustrate the challenges of the modern memory wall on today's GPUs, we profile the H100 using: (1) low batch LLM inference for end-to-end behavior, and (2) standalone dense-linear kernels to isolate bottlenecks.

\textit{\textbf{Experimental Setup:}}
We use NVIDIA’s NVML~\cite{noauthor_nvidia_2024} to measure power on an H100 GPU running CUDA software stacks optimized for HBM3e.
For full-model profiling, we profile Llama-70B with FP8 weights~\cite{noauthor_neuralmagic_2025} on 4×H100s using vLLM~\cite{kwon_efficient_2023} and NVIDIA Dynamo~\cite{elmeleegy_introducing_2025}, running batch-32 inference with 16k prefill and 2k decode tokens (Figure~\ref{fig:h100-characterization}, left).
To isolate bottlenecks, we also benchmark representative dense-linear kernels compiled with PyTorch 2.2~\cite{noauthor_introduction_2023}.

\textbf{\textit{Low Power Efficiency at Low Batch Sizes:}}
Figure~\ref{fig:h100-characterization} (left) shows that during LLM prefill, the H100 uses 90\% of its TDP and achieves high compute utilization. 
In contrast, the decode-phase only uses 34\% of its TDP. 
This observation is reinforced by isolated dense-linear profiling in Figure~\ref{fig:isolated-kernel-profiling} (left); batch sizes $\leq$64 consistently yield $<$30\% TDP. 
The inability to fully utilize power suggests a critical mismatch between the H100 design and low-latency inference.

\textbf{\textit{Low Energy Efficiency at Low Batch Sizes:}}
Figure~\ref{fig:isolated-kernel-profiling} (right) shows that while high AI, compute-bound kernels are energy efficient ($\sim$1.0 pJ/BF16 FLOP), this degrades by 10-1000× for low-batch inference due to non-amortized data movement costs. 
HBM3e accesses alone account for 30-50\% of total energy~\cite{moon_advanced_2023}, with additional losses from on-chip data movement across the H100’s large monolithic die~\cite{choquette_nvidia_2023}. 
The UMA memory system and randomized address mapping further inflate energy costs by 
physically increasing the distance data travels from memory to compute.


\textbf{\textit{Inference Does Not Achieve Peak Memory-BW Utilization: }} 
Our profiling shows
that the H100 only utilizes 32\% of its peak memory bandwidth during distributed LLM decode. 
This observation is consistent with prior work on low-batch inference~\cite{prabhakar_sambanova_2024, zhang_llmcompass_2024, kundu_performance_2024} as well as NVIDIA self-reported benchmarks (20k/2k at 911 OTPS)~\cite{noauthor_nvidia_2024}.
Isolated experiments in Figure~\ref{fig:h100-characterization} (right) show that full bandwidth is only achieved when the working set exceeds $\sim$1GB, which is far larger than typical LLM matrices.
Model sharding and reduced precision (e.g., 16-bit to 4-bit) further reduces each matrix’s footprint per GPU, leading to even lower bandwidth utilization.

Multiple factors contribute to low memory bandwidth utilization: kernel launch overheads become non-negligible for small kernel sizes~\cite{liu_performance_2025}; long memory access latencies cannot be hidden behind compute; and inefficient vector broadcasts between SMs using shared memory limit throughput between layers. 
Together, these limitations show that the H100 memory system is not optimized for the decode phase of LLMs, underscoring the need for a decode optimized architecture.





\section{Capacity Optimized High Bandwidth Memory} \label{sec:bw-per-cap}

Low-batch token generation latency is fundamentally limited by memory bandwidth. 
For dense models like Llama3, consider the case where the model’s memory footprint (weights and KV\$) fits perfectly within the systems memory capacity (100\% capacity utilization). 
In this configuration, all memory capacity is actively used, and token generation latency is determined solely by how quickly that memory can be read. 
This scenario exposes a fundamental constraint: when memory is fully utilized, the minimum achievable latency is set by the ratio of bandwidth to capacity (BW/Cap). 
As a result, BW/Cap emerges as a key metric for evaluating and designing memory systems for bandwidth-bound inference.
Higher BW/Cap enables faster access to the entire model, reducing latency and improving memory efficiency.

Modern memory systems fall short of the bandwidth-to-capacity ratios desirable for efficient low-latency inference.
For example, achieving a 1ms token latency while fully utilizing memory would require a BW/Cap of approximately 1000, which is equivalent to 1TB/s per GB.
In contrast, high-end memory technologies like HBM3e offer much lower BW/Cap ratios. For instance, a single HBM3e stack provides 1280GB/s of bandwidth and 48GB of capacity, yielding a BW/Cap of 27~\cite{lee_134_2024}.
To meet bandwidth targets, system designers must aggregate multiple stacks, which increases total memory capacity far beyond what the model requires and results in severe capacity underutilization.
The fraction of memory actually used is proportional to the ratio between available and required BW/Cap.
In this example, with a target of 1000 and available BW/Cap of 27, only 2.7\% of capacity is effectively utilized.

This mismatch between the desired bandwidth and practical memory capacity defines the memory overprovisioning paradox: High-capacity DRAM-based memories like HBM, GDDR, and LP-DDR \textbf{\textit{buy bandwidth via capacity}} -- scaling and distributing weights across multiple memory modules to increase memory bandwidth, resulting in under-utilized capacity. 
Conversely, SRAM-based architectures \textbf{\textit{buy capacity via bandwidth}} -- offering extreme bandwidth, but struggling to fully utilize it due to excessive sharding across devices caused by limited storage density.

\begin{figure}
    \centering
    \includegraphics[width=\linewidth]{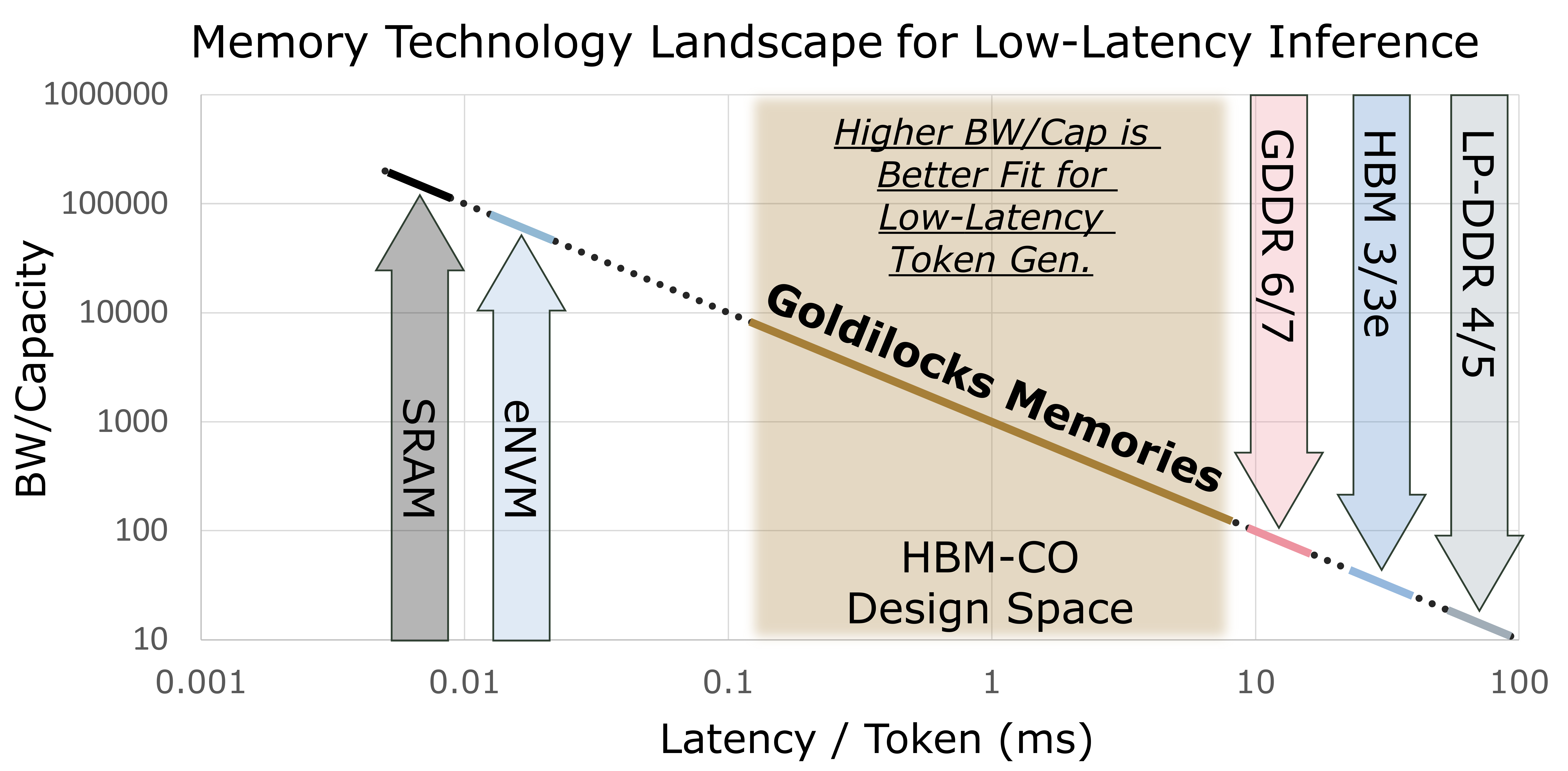}
    \caption{
    Memory technology landscape comparing bandwidth per capacity versus latency per token with 100\% capacity utilization for dense LLMs. A technology gap exists in the \textit{Goldilocks} range for low-latency inference. 
    }
    \label{fig:bw-per-cap}
\end{figure}

Figure~\ref{fig:bw-per-cap} illustrates this design gap: no commercial memory technology occupies the high BW/Cap regime desirable for low-latency LLM inference.

Overprovisioned capacity also introduces energy and cost inefficiencies. 
Prior work~\cite{oconnor_fine-grained_2017} shows that 74\% of HBM energy in streaming workloads is spent on internal data movement, with only 14\% and 12\% attributed to I/O and row activation. 
As capacity increases, internal wire lengths grow, raising energy per bit and reducing efficiency. 
In addition, memory cost scales with capacity due to more silicon area.

To address these challenges, memory capacity per device should become a tunable architectural parameter. 
LLMs differ widely in model size, sparsity, deployment context, and system constraints. 
Each use case has a different optimal BW/Cap profile, often beyond what current technologies can deliver. 
Decoupling capacity from bandwidth would allow system designers to provision memory precisely for application needs, improving performance, efficiency, and cost-effectiveness.

\begin{figure}
    \centering
    \includegraphics[width=.98\linewidth]{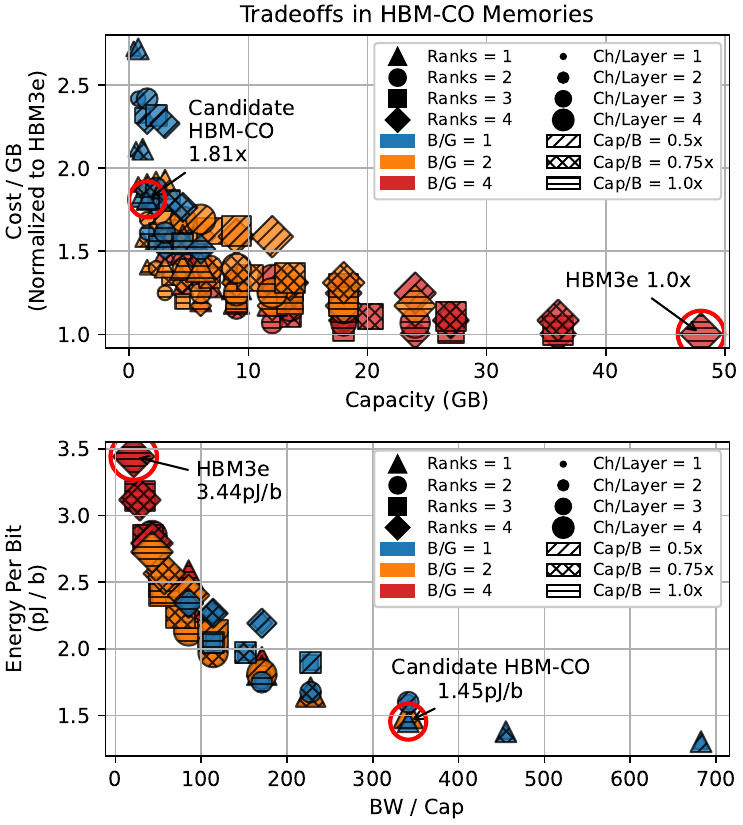}
    \caption{Tradeoffs in \mem{} memories, illustrating that high-BW/Cap memories are up to $\sim$2.5x more energy efficient than an HBM3e device, but $\sim$1.8x the higher cost per GB.}
    \label{fig:hbm-co-dse}
\end{figure}

\textbf{\textit{The Design Space of \mem{} Memories: }} 
To fill the memory technology gap for low-latency inference, we propose a new class of memory devices: \textit{Capacity-Optimized High-Bandwidth Memories} (\mem{}).
We analyzed the HBM architecture and identified  parameters that impact a stacked memory's bandwidth-to-capacity ratio~\cite{chatterjee_architecting_2017, kim_case_2012, kim_present_2024}. 

\textit{HBM Ranks and Layers:}
Each rank consists of four stacked DRAM layers (dies). 
All layers in a rank contribute to higher memory bandwidth, each with separate channels.
However, increasing the number of ranks adds memory capacity but does not increase bandwidth, since the interface is shared. 

\textit{HBM Channels and Pseudo-Channels:}
A DRAM layer is partitioned into four channels, each of which is further split into two pseudo-channels (pCHs) for a total of eight pCHs per layer. 
The 8 pCH across 4 layers per rank fully saturate the memory bandwidth broken into a 32-pCh x 32b IO interface.

\textit{HBM Bank Groups, Banks, and Sub-Arrays:}
\label{bank-group-architecture}
A pseudo-channel contains four bank groups, each with four banks. 
To sustain the full 32 GB/s bandwidth per pCH, only one active bank per bank group is needed
using innovations such as sub-array level parallelism~\cite{kim_exploiting_2018}. 
Four active bank groups per pCH are pipelined to delivers 256 bits per 1 GHz.
Banks are composed of subarrays, which contribute to total capacity but do not impact bandwidth.

\textit{Key Insight to Change the BW/Cap of HBM:}
HBMs achieve peak bandwidth per shoreline with just one active bank per bank group per pseudo-channel.
This means capacity structures such as sub-arrays per bank, banks per bank group, and ranks can be parameterized without changing bandwidth. 

\textbf{\textit{Modeling Energy and Cost for \mem{}: }}
We developed an analytical \mem{} model to capture tradeoffs in bandwidth, capacity, energy, and cost. 
Energy per bit was broken into four components: 
(1) Row Activation: 0.18pJ/bit for streaming workloads~\cite{oconnor_fine-grained_2017, chatterjee_architecting_2017}.
We conservatively model \mem{} with HBM3 timing and activation energy, leaving potential bandwidth and energy gains from its smaller core-die and sub-arrays for a future physical design study.
(2) Data Movement: 0.2pJ/bit/mm, estimated from intra-die routing distances derived from HBM core-die floorplans~\cite{lee_134_2024, ryu_16_2023, park_192-gb_2023}.
(3) TSV Traversal: 0.148pJ/bit/layer, based on 0.8pF TSV capacitance and switching energy~\cite{kim_energy-efficient_2023}. 
(4) I/O Interface: 0.25pJ/bit, drawn from UCIe specs and HBM3e datasheets~\cite{das_sharma_universal_2022, moon_advanced_2023}.
Cost is normalized against an HBM3e  baseline~\cite{patel_ai_2023, zhang_llmcompass_2024}, scaling against silicon area and accounting for non-amortized costs such as base-die logic and TSV footprint. 
At lower capacities, these fixed costs dominate, impacting cost per GB more significantly.
We validate our \mem{} model against HBM3e~\cite{moon_advanced_2023} reported 3.44pJ/bit.

\begin{figure*}
    \centering
    \includegraphics[width=\linewidth]{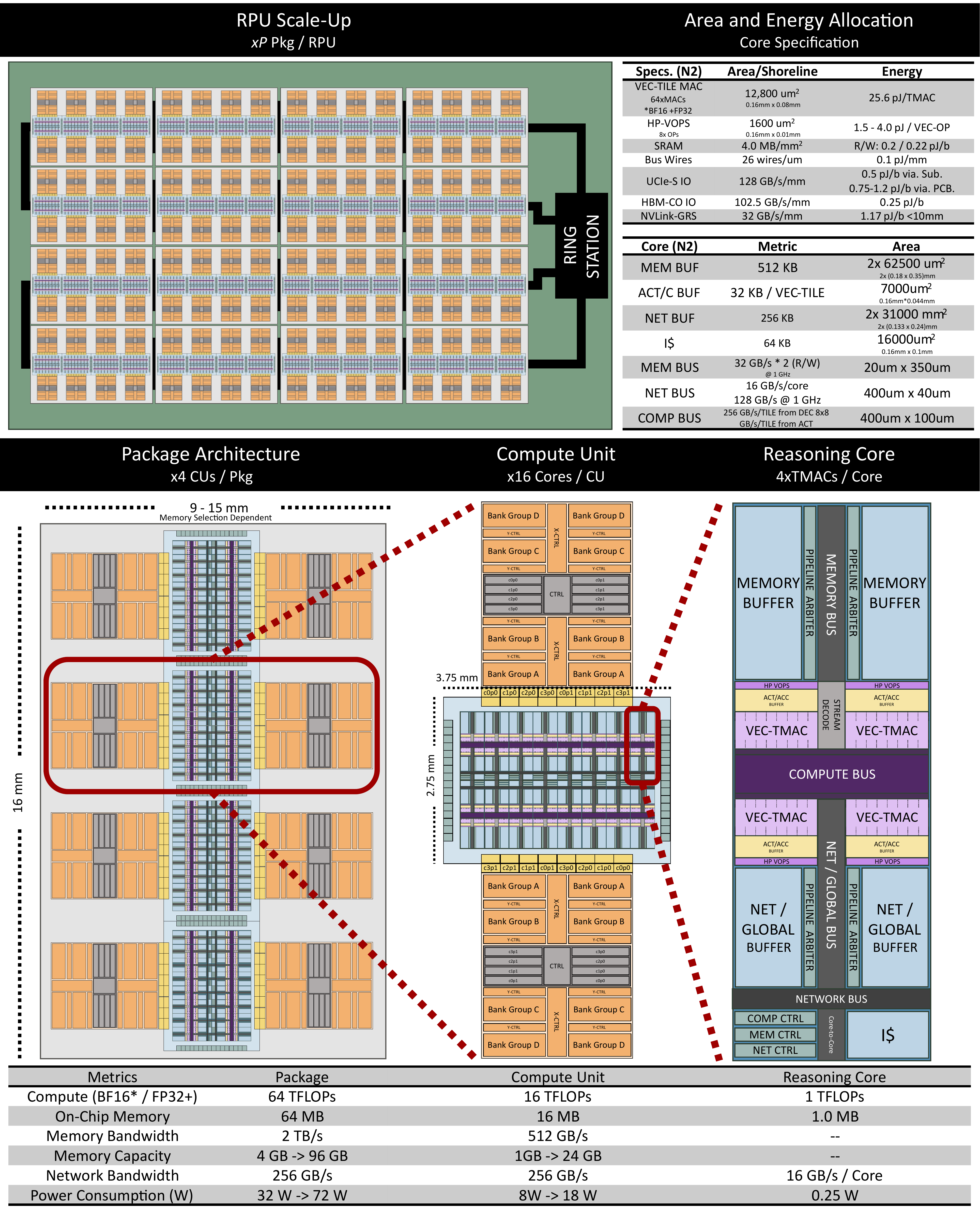}
    \caption{Proposed RPU system architecture for low-latency LLM inference, featuring a chiplet-based design that rebalances compute, memory, and network resources. Each level of the hierarchy, from core micro-architecture to compute units, packages, and ring-station scale-up, is co-optimized for energy-efficient, cost-effective, and scalable memory bandwidth.}
    \label{fig:system-architecture}
\end{figure*}

\textbf{\textit{Design Space Takeaways: }}
Figure~\ref{fig:hbm-co-dse} visualizes the tradeoffs in energy, cost, and BW/Cap for \mem{} memories. 
A candidate Pareto-optimal \mem{} memory has 768MB capacity, 256GB/s bandwidth (BW/Cap = 341), and 1.45pJ/bit energy.
This device offers 2.4× lower energy per bit than HBM3e while maintaining the same bandwidth per shoreline (GB/s/mm). 
This candidate memory BW/Cap leads to an ideal token latency of $2.9$ms per token, falling in the middle of the \textit{Goldilocks} memory range of Figure~\ref{fig:bw-per-cap}. 
An HBM3e system with the same performance would only utilize 7.9\% of its capacity for inference with a dense LLM.

\begin{figure*}
    \centering
    \includegraphics[width=\linewidth]{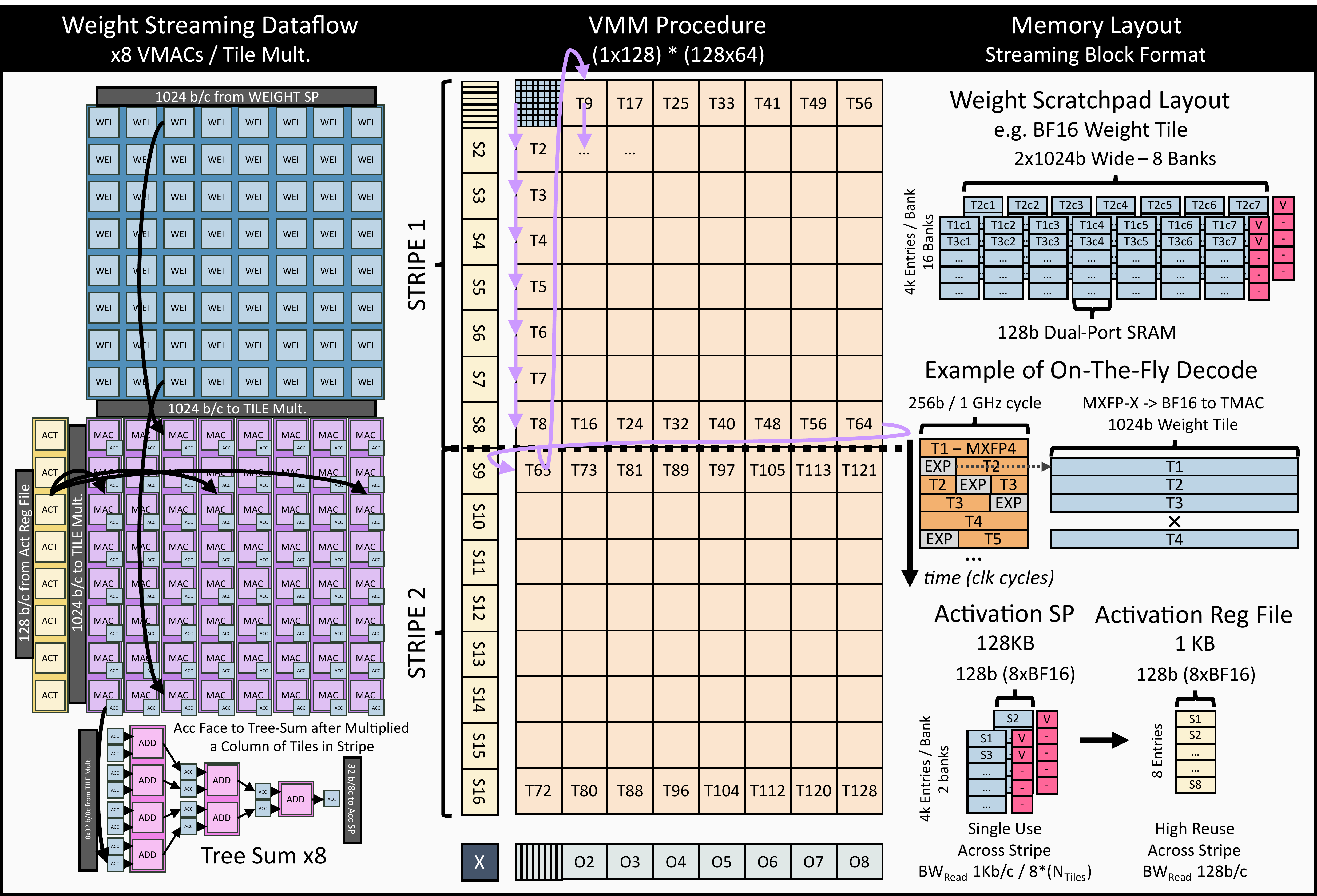}
    \caption{Vector-Tile weight streaming dataflow and VMM procedure following a stripe-based execution. 
    Arrows in \textit{Weight Streaming Dataflow} indicate how activations and weights are moved into TMAC unit -- activations are broadcast across columns while weights are element-wise moved. 
    The arrows in the \textit{VMM Procedure} indicates the order tiles are processed -- column-wise first until eight rows of tiles are processed, then the next column starts, proceeding until all the columns in a stripe are completed. }
    \label{fig:vmm-procedure}
\end{figure*}

This efficiency comes at a cost. 
The candidate is 1.81× more expensive per GB, seemingly violating the foundational DRAM principle of minimizing cost per bit.  
However, for low-latency inference, bandwidth per dollar is the important design metric. 
By trading 192× capacity and 1.81× higher price per GB, the resulting module is 35× lower cost overall, achieving 5× higher bandwidth per dollar than HBM3e.

\section{Compute Fabric for Low-Latency Inference}\label{sec:comp-fabric} \label{subsec:dist-vmm}

Distributed Vector-Matrix Multiplication (VMM)
is the core operation in LLM token generation. 
Given an input vector $V \in \mathbb{R}^{1 \times K}$ and a weight matrix $W \in  \mathbb{R}^{K \times N}$: $O = V * W, \quad O \in \mathbb{R}^{1 \times N}$.
For low-latency inference, this computation must be parallelized efficiently across devices to maintain fast, per-token response times. 
Prior work~\cite{wang_overlap_2022} has exploited the layered structure of AI models, where each layer's output serves as part of the input to the next. 
Consider a system comprising $C$ number of cores.
Sharding $W$ along its columns ensures that each core computes a disjoint portion of the output vector.
The weight matrix is partitioned such that each core stores $W_i \in \mathbb{R}^{K \times \frac{N}{C}}$ and computes its corresponding output fragment $O_i = V * W_i, \quad O_i \in \mathbb{R}^{1 \times \frac{N}{C}}$ 

Since each core holds a portion of the output vector $O$, which serves as the input to the next layer ($O_i$ becomes $V_i$), it can immediately begin computing on its local fragment for the next layer while simultaneously broadcasting its portion of $V$ to other cores. 
This allows each core to progress with available data while receiving the remaining parts of $V$.
This strategy mirrors Cannon’s algorithm for distributed matrix multiplication, where data movement and computation are interleaved to maximize efficiency.

To further increase parallelism, rows of $W$ ($K$-dimension) can be distributed across $G$ cores in a processing groups. 
Using this approach, each core stores weight shard $W_{j,i} \in \mathbb{R}^{\frac{K}{G} \times \frac{N}{C/G}}$ to compute a partial output $O_{j,i}$, requiring a reduction step to sum the intermediate results $O_i = \sum_{j=1}^{G} O_{j,i}$. 
This reduction will always appear on the compute-network critical path, unlike the prior network-broadcast.

Figure~\ref{fig:system-architecture} illustrates the proposed RPU chiplet-based architecture, designed to accelerate distributed VMM for low-latency inference.
The RPU tightly integrates compute and memory across multiple hierarchy levels -- cores, compute units, packages, and ring stations -- to form a scalable and efficient system architecture.

\textbf{\textit{Compute Unit and Reasoning Core: }}
The Compute Unit (CU) is the fundamental building block of the RPU, providing tightly coupled compute and memory resources.
Each CU is constructed with one compute chiplet and two \mem{} chiplets, connected through advanced packaging such as EMIB \cite{mahajan_embedded_2016} or CoWoS-L \cite{hu_cowos_2023}.
The module provides dual 256 GB/s memory shorelines, delivering consistent bandwidth per interface while offering customizable \mem{} capacity.

The particular \mem{} chiplet visualized in Figure~\ref{fig:system-architecture} is derived from the HBM core-die shown in Figure 2 of \cite{park_192-gb_2023}, compacted by reducing banks per group from four to one, ranks from four to one, channels per layer from four to one, but keeping four layers per rank.
Physically, the design reduces the DRAM array region and channel shoreline proportionally, while the TSV, command, and peripheral logic regions are unscaled, occupying roughly one-third of the total die area.

The compute-to-bandwidth ratio for a CU was determined empirically for low-latency inference using MXFP4 formats.
We found that 32~OPs/Byte maximized utilization (Figure~\ref{fig:roofline}); higher ratios offered little benefit and only increased design complexity, silicon area, and energy cost. 
Thus, each 256~GB/s shoreline requires 8~TOPs of compute throughput.

The 256~GB/s shoreline can easily accommodate 512~MAC units along the same horizontal span while leaving adequate space for routing -- defining the \textit{compute shoreline}.
To reach the target 8~TOPs, we stack 16 rows of MACs, organized into 8 reasoning cores. 
Each reasoning core comprises four 8x8 tile-multipliers (TMACs) and connects to its own \mem{} memory pseudo-channel delivering 32~GB/s of memory bandwidth. 
This vertical stacking keeps routing paths short and avoids congestion along the bandwidth edge.
Using both the top and bottom chip edges doubles the number of cores per CU while preserving a balanced area-to-perimeter ratio.



\textbf{\textit{Package Architecture: }}
Four CUs are integrated onto a single package substrate, each equipped with its pair of dedicated \mem{} memories offering 2 TB/s of memory bandwidth. 
At the package level, compute chiplets form a segment of the \textit{outer ring} hierarchy. 
Vector fragments within a CU are forwarded to neighboring CUs in the same package through energy-efficient, short-reach UCIe interconnects~\cite{das_sharma_universal_2022}. 
To minimize communication latency, each core includes a custom DMA engine optimized for fast inter-chiplet transfers, achieving latencies of $\leq$10 ns per CU-to-CU hop, which is similar to prior works~\cite{shao_simba_2019, zimmer_011_2019, gratz_-chip_2007, campanoni_helix-rc_2014}. 
Each compute chiplet uses a unified UCIe-S physical interface with segmented drivers: in-package links run at low voltage and high frequency (0.5 pJ/bit), while off-package links operate up to 16 GT/s with 0.75-1.2 pJ/bit energy\cite{das_sharma_universal_2022, poulton_117-pjb_2019}, defining the system’s outer-ring bandwidth at 128 GB/s/mm.

\textbf{\textit{RPU Scale-Up: }}
Multiple packages are soldered onto a PCB to form the outer ring topology, connected via a Ring Station. 
Communication between packages leverages PCB-routed interconnects, designed specifically for short-reach ($<$10 mm) data transfers.
A secondary purpose of the Ring-Station is to network outside the system (e.g., 100Gb Ethernet).

An RPU is defined as a scalable compute system, composed of multiple co-packaged CUs, assembled on a board.
Similar to how GPUs scale across datacenter and edge deployments by varying the number of CUDA cores, RPUs scale by composing different numbers of CUs.
Our modular architecture enables flexible configurations to meet diverse performance, capacity, energy, and cost targets.

\section{Micro-Architecture}\label{sec:micro-architecture}

\textbf{\textit{NUMA Domains and Data Dependent Synchronization: }}
A central design principle of our microarchitecture is a fully NUMA-based system.
Each compute core within a CU forms an independent NUMA domain, without shared memory between cores. 
All data movement across domains is explicitly managed via software-programmable DMA engines and data-dependenct synchronization.
This eliminates coherence overhead, enables deterministic execution, and ensures scalable performance for dataflow-dominated workloads like LLM inference.
Thus, the RPU favors \textit{bespoke datapaths over generalized programming models.}

\textit{NUMA at All Scales: }
Each core includes three programmable data pipelines, each operating within its local NUMA boundary.
The Memory DMA transfers data between the core’s dedicated \mem{} memory channel and its memory buffer. 
The Compute DMA reads from memory or network buffers and feeds data into the compute pipeline.
The Network DMAs manage all inter-core and inter-chiplet communication, linking each core to neighboring cores within a CU and the positionally aligned core in adjacent CUs. Incoming data is written to the network buffer and may be consumed locally and/or forwarded using custom forwarding instructions. This supports efficient collectives and data reuse across chiplets.

\textbf{\textit{Pipeline Arbiters: }}
We developed Pipeline Arbiters to synchronize decoupled memory, compute, and network pipelines. 
These lightweight, software-managed mechanisms are embedded within each core’s SRAM buffer.
Each SRAM buffer entry includes a 2-bit valid counter that tracks the expected number of asynchronous consumers. 
DMA operations are programmed with a \textit{valid\_count} when writing and may optionally enable a \textit{check\_valid} flag to stall if the target address is occupied.
On the read side, consumers can use \textit{check\_valid} to stall until data is ready and optionally decrement the valid counter after access. 
For example, a Network DMA may set \textit{valid\_count=2} since activations will be consumed by (1) the compute pipeline and (2) asynchronously forwarded to neighboring cores.

To guarantee mutual exclusion, each buffer entry is accessed through a hardware-enforced arbitration mechanism that serializes requests from multiple consumers. 
Accesses are prioritized using a software-configurable policy, ensuring that only one DMA engine can read, write, or update the valid counter at a time. 
This enforces atomicity at the buffer-entry level and prevents race conditions across the memory, compute, and network pipelines.
By managing synchronization through software-defined counters and flags, Pipeline Arbiters enable fine-grained, data-driven execution between NUMA domains with blocking and non-blocking semantics.

\textbf{\textit{TMAC and HP-VOPs: }}
The vector-tile MAC (TMAC) is the core computational unit for accelerating the VMM kernel, as shown in Figure~\ref{fig:vmm-procedure}. 
Each TMAC consists of 64 MAC units arranged in an 8×8 array, performing BF16 multiplies with FP32 accumulations. 
This structure allows one activation vector to be broadcast across 8 columns of the weight matrix, computing 64 MACs per cycle using a weight-streaming, output-stationary dataflow.

To maximize on-chip reuse of activation data and minimize accumulation write-back pressure, the VMM algorithm is organized into stripes. 
A stripe is a groups of 8 vertically stacked tiles spanning all columns of the weight shard. 
Activation shards per stripe contains 64 BF16 values, stored in a dedicated register file close to the tile multipliers. 
These values are initially fetched from the network buffer, then reused across all tile columns before being retired.

The tile multipliers first iterate over the tile-rows within a stripe. 
After processing a column of tiles, the accumulated face is reduced via a column-wise (3-stage) tree sum. 
These results are written back to a local register file to be read back for the next stripe, leveraging the fact that each core typically operates on small output shards ($<$256 elements) in highly distributed VMMs. 
Once all the weight matrix columns of a stripe are computed, the next activation stripe shard is loaded from the network buffer, and the process repeats.

This striping approach is essential for three reasons:
(1) Traversing columns first (inner-product style) would require the full activation vector to be stored on-chip, stalling compute during the vector broadcast across all CUs.
(2) Traversing rows first (outer-product style) would result in high writeback bandwidth due to frequent partial sum updates.
(3) By processing one stripe at a time, we minimize on-chip bandwidth requirements and enable fine-grained overlap of computation and communication; the next  activation shard is collected in the network buffer while compute works on the current shard.

In addition to the tile multipliers, each core includes a general-purpose, high-precision (FP32) vector operations (HP-VOPs) accelerator, enabling support for key functions in LLM workloads (e.g., SiLU, GeLU, normalization, and rotary embeddings). 
Because overall performance is dominated by memory bandwidth, we can afford to allocate area to high-precision computation without significant impact on energy or latency. 
This enables numerical accuracy, particularly important for operations sensitive to precision loss such as attention.

\textbf{\textit{Stream Decoder: }}
To reduce latency and storage overhead, weight tiles are stored in compressed block formats in memory and transferred on-chip to the memory buffer by the memory DMA engine. 
Next, the compute DMA streams compressed weights into the \textit{Stream Decoder}, which performs on-the-fly dequantization, converting block-quantized values into standard BF16. 
This continues until a full batch of 64 BF16 values is reconstructed, corresponding to a single weight tile.
Once dequantized, the tile is broadcast across all active tile multipliers within the core via a 1024-bit wide compute bus.

Our stream decoder supports on-the-fly dequantization of multiple formats, including BFP~\cite{rouhani_pushing_2020}, MxFP~\cite{darvish_rouhani_shared_2023}, and NxFP~\cite{lo_nanoscaling_2024}, with configurable bitwidths ranging from 4 to 8 bits. 
This flexibility allows us to efficiently compress weights off-chip while preserving the ability to compute at full precision on-chip, minimizing off-chip capacity and energy without compromising accuracy.




\begin{figure*}
    \centering
    \includegraphics[width=\linewidth]{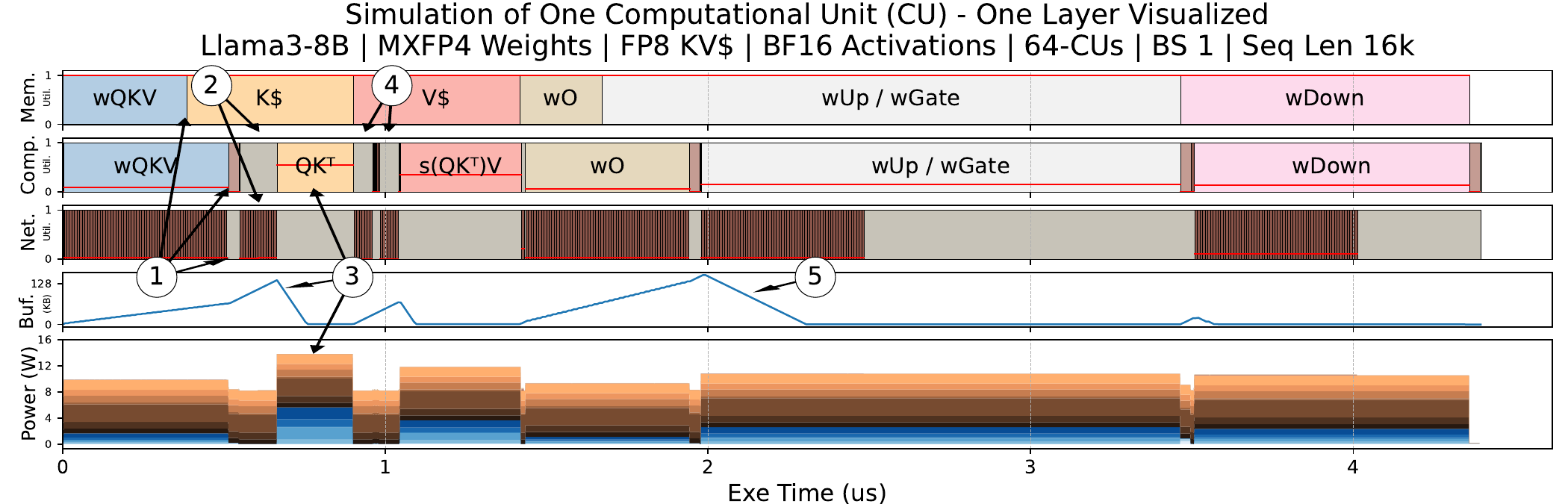}\\
    \medskip
    \includegraphics[width=\linewidth]{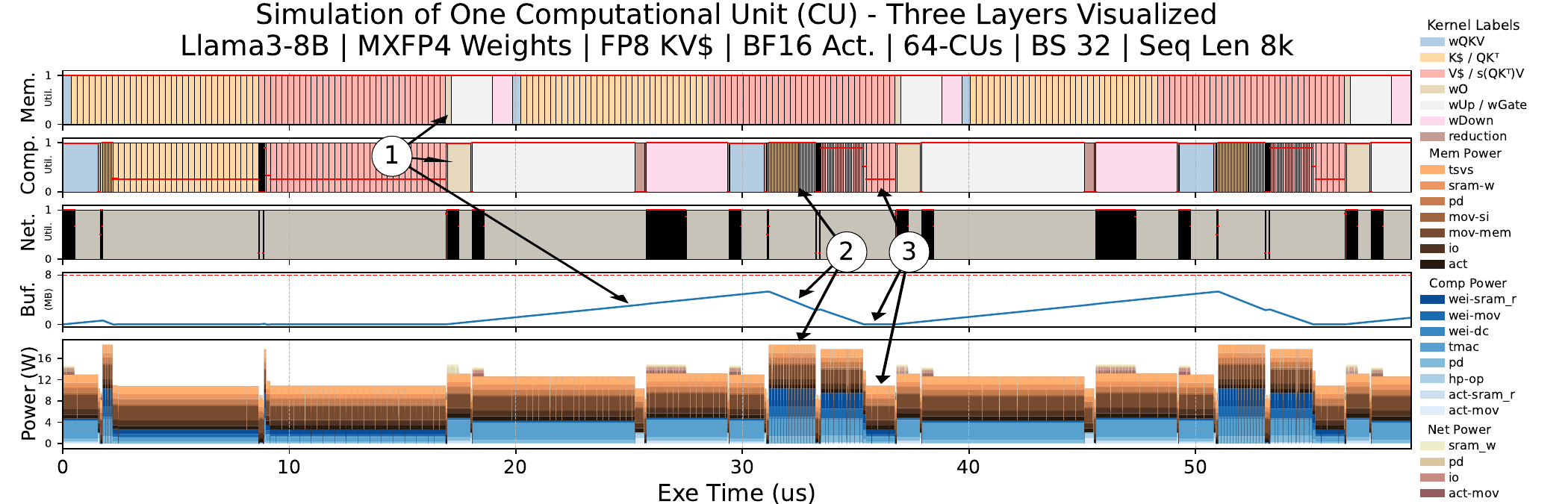}
    \caption{Simulation of one CU in a 64-CU system running Llama3-8B. 
    Top: Batch size 1, seq len 16k. Bottom: Batch size 32, seq len 8k. 
    Each timeline shows memory, compute, and network utilization, buffer usage, and power. Memory power dominates total system power, highlighting the RPU’s unique power provisioning. 
    The plot also shows that batch size 32 generates tokens $\sim$13× slower than batch size 1, primarily due to sequential KV\$ computations.}
    \label{fig:simulation}
\end{figure*}

\section{RPU Software Stack and Simulation}\label{sec:simulation}

\textbf{\textit{RPU ISA and Compiler: }} The RPU software stack provides a deterministic compilation flow from PyTorch graphs to hardware execution.
The RPU ISA hardens optimized vector-matrix and elementwise dataflows directly into hardware, exposing each as a single CISC-style instruction.
Each instruction specifies operand addresses, tensor dimensions, and data types, while the hardware executes a fixed streaming schedule with near-roofline utilization.
Computation follows a \textit{push-based} dataflow: DMA engines deterministically inject data into buffers, and pipelines advance when inputs are ready.
Hardware-managed pipeline arbiter flags are embedded within each instruction to synchronize compute, memory, and network pipelines, eliminating the need for software polling and ensuring deadlock-free progress.

A lightweight Python compiler traces PyTorch operations and lowers them to RPU primitives.
For example, a \texttt{torch.nn.Linear} layer compiles into a three-stage microkernel -- Loading, Looping, and Launching -- that programs DMAs, drives the VMM pipeline, and forwards activations, respectively.
The compiler statically orders all DMA and compute instructions, pre-shards and quantizes weights, and generates synchronized instruction streams for the memory, compute, and network pipelines.
Together, the ISA and compiler form a compact, deterministic toolchain that provides predictable, near-roofline performance for token generation.


\textbf{\textit{Deployment: }}
Each RPU core includes a lightweight instruction-fetch pipeline that executes a small set of long-running instructions for a full LLM. 
This enables fully autonomous execution, eliminating the host-driven offload model used by GPUs. 
The host processor performs only coordination tasks such as transferring KV\$ from the prefill engine into RPU memory. 
After each transformer layer, the RPU triggers an interrupt to the host and reports generated tokens or completed queries and returns the corresponding KV caches.

\textbf{\textit{RTL Simulations: }}
We implement a proof-of-concept RPU in SystemC using Catapult HLS~\cite{tambe_edgebert_2021, zhang_camel_2024}, targeting TSMC N16 and projecting to N2 using published scaling factors~\cite{shilov_tsmc_2018, frumusanu_tsmc_2020, shilov_tsmcs_2024}. 
RTL simulation includes a single-core CU, multi-CU packaging, and board-level integration. 
Key microkernels (e.g., VMM, DMA) are synthesized using VCS, Design Compiler, and PowerPro to extract calibrated energy and area. 
SRAM and interconnect energy are modeled analytically, and memory energy is provided by our HBM-CO analytical model, which captures the energy and bandwidth tradeoffs of capacity-optimized memory devices.

Compiled PyTorch transformer layers were executed on the RTL model to verify functional correctness and dataflow behavior. 
However, full-system cycle-accurate simulation remains computationally intensive: simulating a \texttt{2k~×~2k} VMM on a 4-core RPU requires approximately 6.5 minutes, making end-to-end design space exploration (DSE) for LLM inference impractical at the RTL level.

\textbf{\textit{Event Driven Simulation: }} 
To address RTL simulation latency, we developed a higher-level event-driven simulator that reproduces the behavior of the RTL model using symbolic transactions that capture address, size, and type instead of real tensor data.
The simulator models all key microarchitectural events -- data transfers, stalls, and arbitration -- using parameters calibrated to RTL throughput, bandwidth, and latency.

The event-driven simulator runs orders of magnitude faster than RTL while matching its latency and power estimates.
It supports full-model DSE across models, batch sizes, sequence lengths, memory devices, and scales of deployments, while exposing transient behaviors such as buffer occupancy, pipeline stalls, and synchronization delays (Figure~\ref{fig:simulation}). 
The simulator also serves as a debugging and validation tool: violations of data dependencies appear as execution stalls, allowing developers to visually trace and correct synchronization issues in the simulation framework prior to deployment.

\textbf{\textit{Simulation Results: }}
Figure~\ref{fig:simulation} shows a simulation of the first transformer layers in Llama3-8B on a 64-CU RPU using the candidate \mem{} from Section~\ref{sec:bw-per-cap}. 
We compare batch sizes 1 and 32, breaking down execution across memory, compute, and network pipelines, along with buffer and power traces. Each row represents a single CU, with red lines indicating average utilization per kernel.
We highlight instances where the RPU's decoupled pipelines enable out-of-order execution between memory, compute, and communication to unlock behaviors that conventional architectures cannot exploit.

\textbf{\textit{Batch Size 1: }}
\circled{1}
During the first VMM (\textit{wQKV}) execution is bounded by network-latency to broadcast activations across all CUs.
This is reflected in the timeline by low network bandwidth utilization, and early memory-pipeline completion. 
Traditional architectures would stall the memory; however, the decoupled memory pipeline enables the RPU to simply continue fetching weights from memory to the on-chip buffer, while the compute pipeline lags the memory stream by $\sim$80kB waiting for activations to arrive from the network. 
During this period, power consumption is dominated by reading weights from memory: $\sim$6.7W at full BW / CU (\textit{512 GB/sec}) and $\sim$1.7pJ/b datapath to write to the memory-buffer.

\circled{2} 
Prior to the \textit{QK$^\top$} computation, compute stalls as each CU gathers its shard of the \textit{Q}, \textit{K}, and \textit{V} vectors -- 32 \textit{Q} heads and 8 \textit{KV} heads distributed across 64 CUs means each \textit{Q}-vector spans two CUs while \textit{KV}-vectors span eight CUs. 
Similarly, during \circled{4}, the compute stalls waiting first for a distributed \textit{max} collective to calculate \textit{softmax}, then an \textit{exp-sum} reduction across the CUs sharing the 8 GQA-heads. 
These examples of cross-CU synchronization and network delays stall the compute pipeline, while the memory pipeline continues to prefetch and move weights and \textit{KV\$} to the on-chip buffer.
These types of network latency-bound periods are common in tensor-parallel distributed systems, where latency-bound network collectives are often on the orders of $\mu$seconds, leading to periods of fully stalled execution. 
The RPU allows the memory pipeline to continue prefetching ahead of the compute stream, effectively eliminating any incurred network collective overheads.

\circled{3} 
During the \textit{QK$^\top$} computation, each \textit{Q}-vector (4 per GQA head) is assigned to a TMAC, while prefetched \textit{K\$} entries are broadcast from the memory buffer. 
Similarly, in \circled{5} the \textit{wUp/wGate} phase drives compute to full utilization while processing the accumulated weights.
This is a unique opportunistic moment for the RPU, as decoupled pipelines enabled the memory-bandwidth to stay fully saturated throughout network delays and compute stalls. 
Later, this enables the compute to play ``catch-up'' until all the available data is consumed from the on-chip buffer, returning to the memory-bandwidth bound performance. 
Power during this period rises due to higher compute utilization, from $\sim$1.5W to $\sim$5W per CU, reflecting full datapath activity. 
Once the buffer is drained, compute utilization returns to the memory-bound performance.


\textbf{\textit{Batch Size 32: }}
\circled{1}
With a larger batch size, weight matrix operations become compute-bound.
Specifically, during \textit{wUp/wGate}, weight are read from memory in $\sim$2$\mu$s, while compute is $\sim$4x longer.
While the RPU is compute-bound operating on the \textit{wUp/wGate} computation, the memory pipeline prefetches ahead, streaming in \textit{KV\$} and filling each CU’s buffer with $\sim$6MB of weights totaling $\sim$384MB system-wide. 
This lookahead window is beyond the capabilities of GPU architectures like H100, which lack both the on-chip capacity and pipeline decoupling to absorb such deep prefetching. 

\circled{2}
After a sequence of compute-bound weight matrix multiplications, the attention computation begins, entering a \textit{KV\$}-intensive phase.
Unlike weights, \textit{KV\$} entries are query-unique, offering reuse only among GQA heads. 
Thus, this phase is inherently memory-bandwidth-bound.
However, \textit{KV\$} has already been prefetched on-chip which allows the system to stream \textit{KV\$} directly from the on-chip buffer and operate at a compute-bound performance.
As a result, the buffer drains rapidly, as the compute ``catches-up'' to the memory stream.

A batched LLM decode system naturally alternates between compute-bound weight layers and memory-bound \textit{KV\$} layers, leading to pipeline underutilization on traditional architectures.
Our design breaks this limitation by \textbf{\textit{smoothing out the bimodal workload}} (compute-bound weights vs memory-bound \textit{KV\$}) and \textbf{\textit{absorbing phase-imbalance into the memory buffer}}, letting decoupled pipelines handle them independently.
By smoothing performance across layers, we enable sustained utilization of both memory and compute.
Without this buffering strategy, overall latency would increase by up to 1.6×.


\circled{3}
Once the buffer is drained, compute returns to the memory-bound performance and power drop accordingly, marking the end of the amortization window.
Larger batch sizes allow deeper prefetching, leading to a tradeoff in sequence length versus batch size to fully saturate bandwidth.



\begin{figure}
    \centering
    \includegraphics[width=\linewidth]{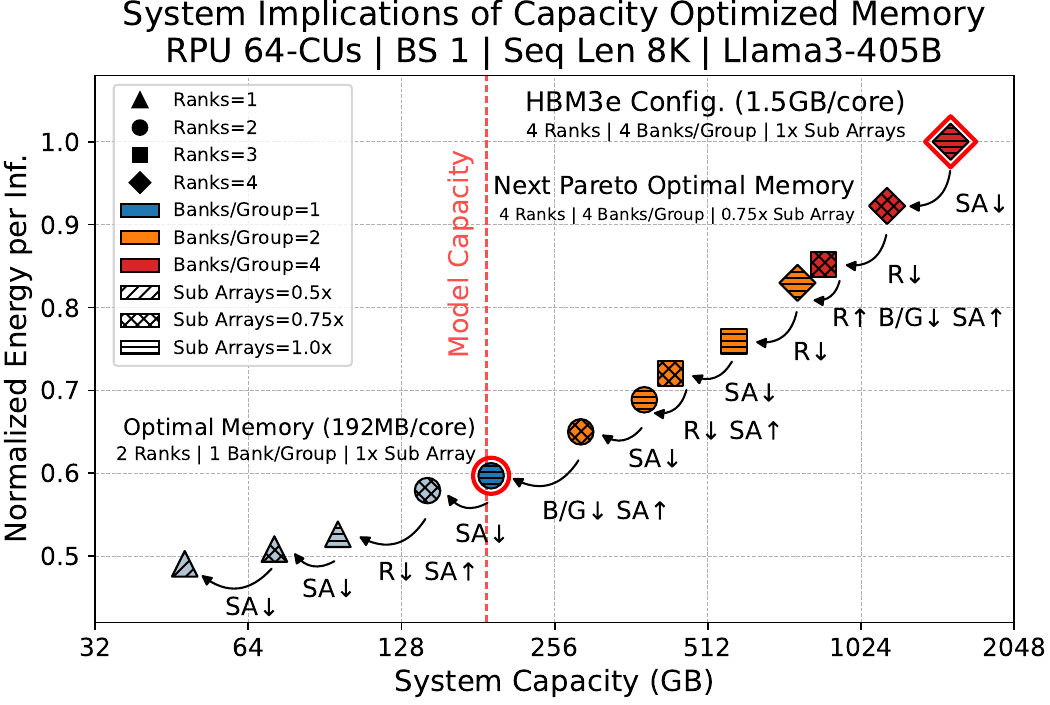}
    \caption{Pareto frontier of \mem{} memories for Llama3-405B inference on 64-CU RPU, annotated by stepwise changes in optimal \mem{}. 
    These represent the set of \mem{} chiplets useful for a memory-chiplet ecosystem.}
    \label{fig:hbm-co-sys-implication}
\end{figure}

\section{System Implications of \mem{}}\label{sec:hbm-co-system-implications}

\mem{} memories improve efficiency at the device level, but their true benefits emerge only through full-system evaluation.
Figure~\ref{fig:hbm-co-sys-implication} shows energy per inference (y-axis) versus system memory capacity for a 64-CU RPU.
\mem{} memory configurations form a Pareto frontier showing the energy-capacity tradeoff; non-optimal points are omitted for clarity.
Capacity reductions are progressively applied to an HBM3e-like memory to traverse the Pareto frontier.
The best capacity reduction strategy is annotated between configurations. 
The optimal \mem{} has the smallest device capacity that meets the system-level requirement to store the target model.


For a 64-CU RPU running Llama3-405B with a single query and an 8k sequence length, the optimal \mem{} configuration has a memory capacity of 192 MB per core.
Compared to HBM3e, \mem{} reduces the energy per bit by 2× from the memory cell to the IO, while at the system-level the energy per inference improves by 1.7× due to memory dominating the energy consumption. 
A similar tradeoff exists for system cost. 
Despite a 1.6× higher cost per GB, reduced capacity \mem{} yields a 5.2× decrease in per-device cost, translating to a 4.3× total system cost reduction when factoring in compute, interposer, and substrate.


As illustrated in Figure~\ref{fig:hbm-co-sys-implication}, several \mem{} configurations offer even lower energy per inference but remain inaccessible at the current 64-CU scale due to their limited memory capacity.
Unlocking these more energy-efficient memories requires increasing the number of CUs, thereby decreasing the required memory per CU. 

\begin{figure}
    \centering
    \includegraphics[width=\linewidth]{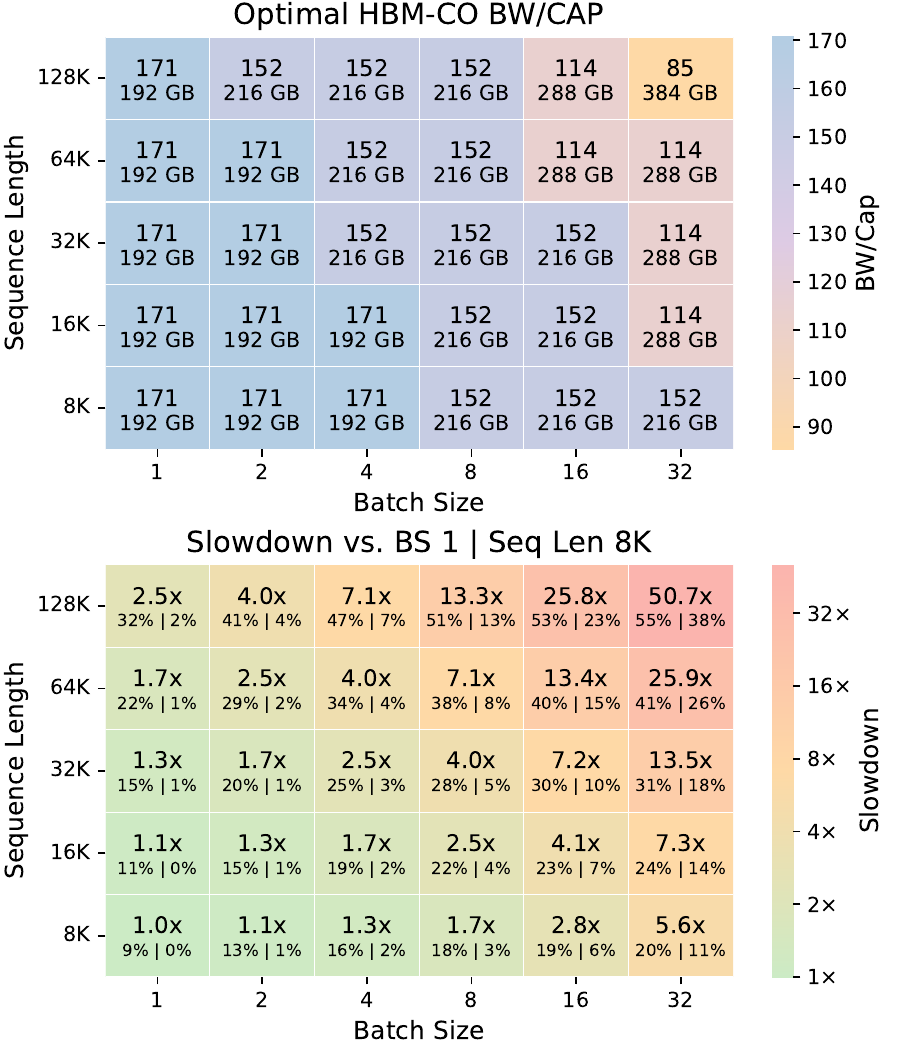}
    \caption{
    RPU with 64 CUs running Llama4-Maverick showing batch size versus sequence length, comparing optimal HBM-CO BW/Cap and slowdown relative to BS=1, Seq Len=8K. Slowdown sub-metrics indicate the fraction of capacity used for KV cache versus active parameters and total capacity.}
    \label{fig:hbm-co-batch-vs-seq}
\end{figure}

A key goal in deploying the RPU is to maintain workload flexibility without proliferating hardware SKUs.
In the emerging chiplet ecosystem, the \mem{} designs along the Pareto frontier of Figure \ref{fig:hbm-co-sys-implication} are sufficient to cover the useful BW/Cap design space.
These chiplets can be mixed and matched at the package level to enable design customizations without fabricating a new ASIC.
Figure~\ref{fig:hbm-co-batch-vs-seq} extends this idea by showing how to select among these variants for a given workload.

Figure~\ref{fig:hbm-co-batch-vs-seq} (top) is an \mem{} SKU selection map for a 64-CU RPU running Llama4-Maverick.
Each memory chiplet has a fixed bandwidth interface, resulting in a total system bandwidth of 32 TB/s.
Given this fixed bandwidth, system capacity is optimized by selecting the most efficient \mem{} chiplet configuration from Figure \ref{fig:hbm-co-sys-implication}, to minimize both energy per inference and overall cost while satisfying capacity for each batch size and sequence length combination.

High BW/Cap SKUs maximize efficiency but limit the range of supported batch and sequence lengths, while lower BW/Cap SKUs trade some efficiency for broader capacity coverage.
Importantly, Figure~\ref{fig:hbm-co-batch-vs-seq} (top) shows that high-BW/Cap memories (5-6x HBM3e) are better suited for long-context, low-batch inference, which underscores the capacity overprovisioning of using off-the-shelf HBM3e. 
Increasing the number of CUs raises the optimal BW/Cap, enhancing efficiency at scale.

Figure~\ref{fig:hbm-co-batch-vs-seq} (bottom) quantifies how batching and sequence length impact latency. 
As batch size or sequence length increase, the per-query token generation latency increases.
This is illustrated by tools such as InferenceMax \cite{semianalysis_inferencemax_nodate}, highlighting that low-batch inference is key for low latency.
Longer sequences also intensify bandwidth pressure during attention -- more than 50\% of the active parameters are KV\$ for BS=8 128k. 
Therefore, the relative efficiency gap between the RPU and conventional GPUs widens due to the RPU’s bandwidth advantage, underscoring its use for long-context, low-latency inference.


\section{Strong Scaling Analysis}\label{sec:strong-scaling}

\begin{figure}
    \centering
    \includegraphics[width=.95\linewidth]{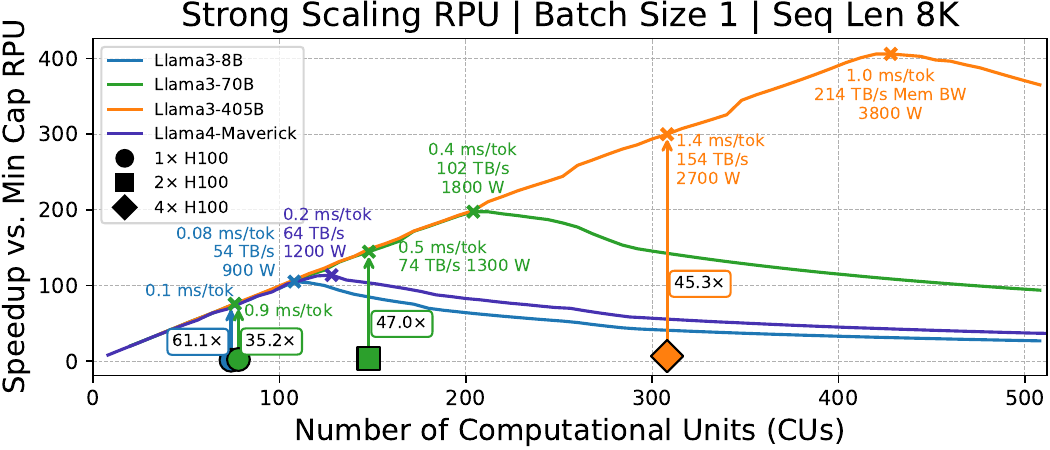}\\
    \medskip
    \includegraphics[width=.95\linewidth]{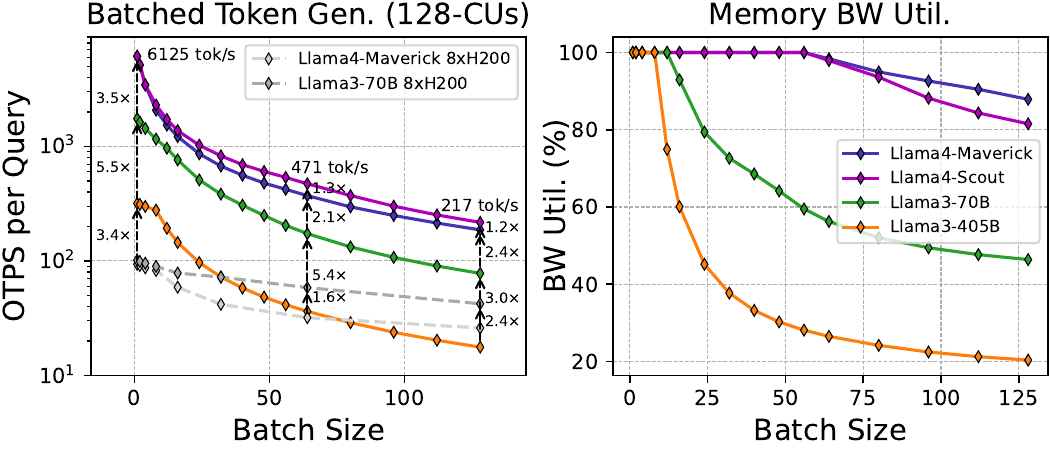}
    
    \caption{Top: Strong scaling for Llama models under ISO TDP vs H100. Bottom: Output tokens per second per query (8xH200 from~\cite{noauthor_hardware_2025}) and bandwidth utilization versus batch size.}
    \label{fig:strong-scaling}
\end{figure}

\textbf{\textit{Strong Scaling Analysis: }}
We conduct a strong-scaling study by varying the number of CUs. 
Speedup is reported relative to the smallest configuration capable of fitting each model. 
Figure~\ref{fig:strong-scaling} shows results for Llama models, compared against an NVIDIA H100 at ISO TDP using the methodology from Section~\ref{sec:motivation} with 4-bit weights and 16-bit activations~\cite{frantar_marlin_2025} and full tensor-parallelism.


\textbf{\textit{Batch Size 1 -- Fastest Thinking Speed: }}
Batch size 1 represents the fastest possible ``thinking speed'' of a model. 
At ISO TDP, the RPU significantly outperforms H100 inference. 
Notably, the RPU latency is 47.0× faster than a 2xH100 at 1400W TDP for Llama3-70B and 45.3× faster than a 4×H100 at 2800W TDP for Llama3-405B.
The 405B example is illustrated in Figure~\ref{fig:strong-scaling} (top) by the orange diamond (4xH100s) aligned to a 308 CU RPU system at ISO TDP. 

Even more compelling is the peak performance of the RPU, achieved by scaling to the optimal number of CUs for each model:
Llama3-70B at 204 CUs achieves 0.4 ms/token, and
Llama3-405B at 428 CUs achieves 1.0 ms/token.
Llama4-Maverick at 128 CUs achieves 0.2 ms/token.
These are the fastest token generation latencies reported to date for these models. 
\textit{Notably, we are the first system capable of sustaining over 200 TB/s of tensor-parallel memory bandwidth during inference for a 405B parameter model.}

Beyond these scales, performance plateaus as broadcasting the activation becomes the bottleneck.
To overcome this limit, we propose two future directions:
1) Reduce on-chip forwarding latency.
2) Reduces hop count by adding another level of scale-out to Figure~\ref{fig:system-architecture} which interconnects ring-stations.

An important insight from this study is that memory customization enables model- and deployment-specific system design.
For instance, an RPU designed for the high-performance edge running Llama3-70B achieves 3.5ms/token using a memory configuration with BW/Cap=227 at 220W TDP. 
Separately, an edge-optimized system for Llama4-Maverick achieves 1.1ms/token at BW/Cap=38 (comparable to HBM3e) and 260W TDP. 
In contrast, for datacenter-scale deployments targeting 1kW TDP, we design an RPU for Llama3-70B with more CUs that achieves 0.65ms/token with a BW/Cap of 682 (the highest in our design space) while a datacenter RPU for Llama4-Maverick reaches 0.24ms/token at BW/Cap=170. 

\begin{figure}
    \centering
    \includegraphics[width=\linewidth]{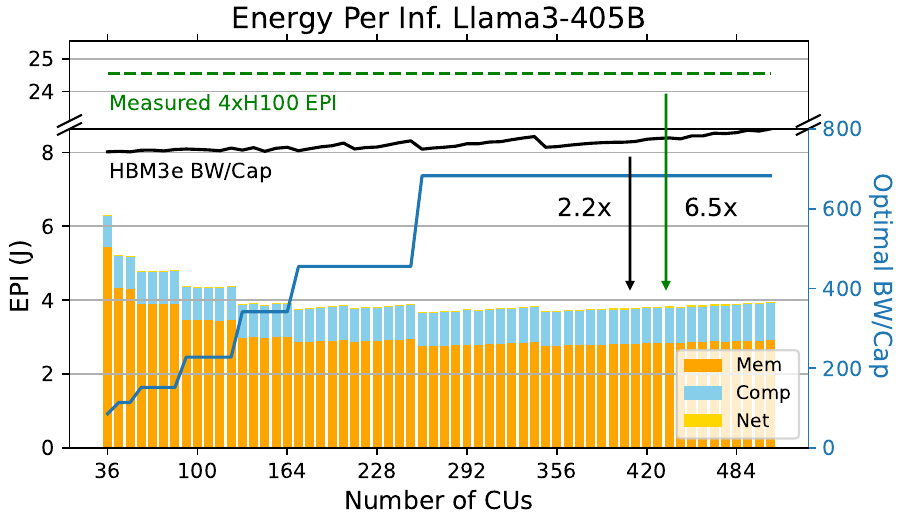}
    \includegraphics[width=\linewidth]{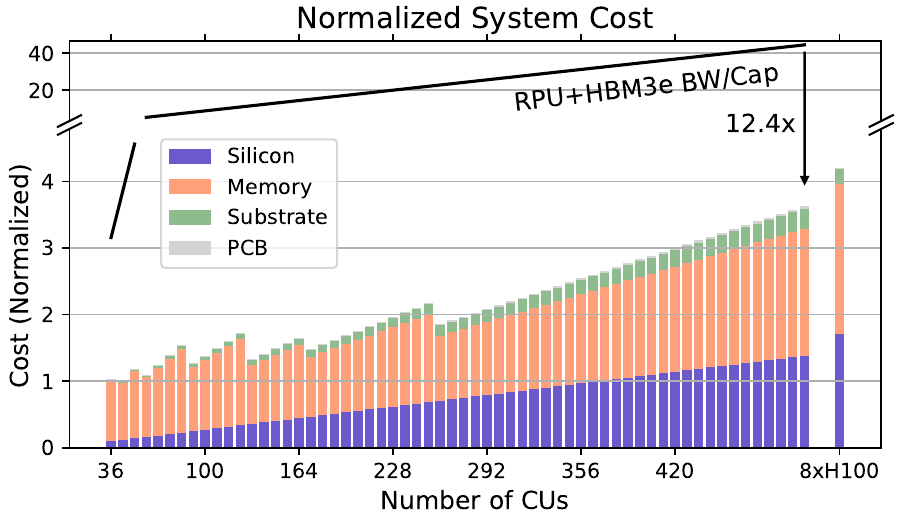}
    \caption{Energy and cost analysis for scales of CUs running Llama3-405B at batch size 1. Top: Energy per inference and optimal memory selection. Bottom: Normalized system cost.}
    \label{fig:energy-per-inf}
\end{figure}

\textbf{\textit{Batched Inference on the RPU:}}
Figure~\ref{fig:strong-scaling} (bottom left) compares output tokens/sec per query across Llama models using a 128 CU RPU and an 8×H200 baseline reported by~\cite{noauthor_hardware_2025}. 
Llama4-Scout achieves the highest throughput across all batch sizes, closely followed by Llama4-Maverick. 
Activating more unique experts in Maverick reduces per-expert parallelism, leading to a 1.2-1.3× decrease in performance compared to Scout’s 16-expert configuration.

As batch size increases, per-query throughput decreases primarily due to serialized KV cache computations. 
Figure~\ref{fig:strong-scaling} (bottom right) demonstrates this behavior, indicating all models operate in a memory bandwidth-bound regime up to a batch size of 8. 
Beyond this point, the Llama3-405B model becomes compute-bound, as its attention mechanism features a high arithmetic intensity (16 queries per KV head), saturating available compute resources.

In contrast, the Llama4 models maintain high memory bandwidth utilization (above 80\%) up to batch size 128. 
Their attention design, with only 5 queries per KV head, and MoE structure, are low per-token arithmetic intensity. 
Our layer smoothing technique balances arithmetic intensity across compute-bound layers (e.g., shared projections and MLP layers) and memory-bound layers (e.g.,  attention and MoE layers). 
Between the two Llama4 variants, Scout becomes compute-bound earlier due to heavier per-expert loads (only 16 experts) versus Maverick’s 128-expert setup distributes tokens across more experts, preserving memory-bandwidth demands.


\textbf{\textit{Energy and Cost Analysis: }}
Figure~\ref{fig:energy-per-inf} (top) extends our strong scaling results by analyzing energy per inference for Llama3-405B at batch size 1.
The majority of the energy is consumed by memory accesses, making memory selection a dominant factor in system efficiency. 
To minimize energy, we explore \mem{} design points along the Pareto frontier (Section~\ref{sec:hbm-co-system-implications}), selecting the BW/Cap ratio that best matches system scale and capacity requirements.

At smaller scales, lower BW/Cap memory modules are required to meet overall capacity needs, resulting in higher energy per inference. 
As the system scale increases, each CU stores a smaller fraction of the model, allowing for a higher BW/Cap memory with lower capacity and improved energy efficiency. 
Energy per inference improves steadily with scale until 268 CUs where the highest BW/Cap memory module in the design space is selected.

Compared to an HBM3e BW/Cap memory, the \mem{} memory improves energy efficiency by up to 2.2×. 
Similarly, compared to a 4×H100 system running Llama3-405B, the \mem{} optimized configuration achieves 6.5× lower energy per inference.
Combined with the latency-optimized design point at 428 CUs, this translates to a 412× improvement in energy-delay product (EDP) relative to the 4×H100 baseline.
This highlights the power of co-designing memory and compute around bandwidth per capacity to unlock both latency and efficiency at scale.

Figure~\ref{fig:energy-per-inf} (bottom) shows the normalized total system cost broken down by silicon, memory, substrate, and PCB. 
Costs are normalized to the smallest valid configuration for Llama3-405B. 
As expected, compute cost grows linearly with CU count, while memory cost increases sublinearly due to adaptive \mem{} selection.
At each scale, the memory configuration is selected from the \mem{} Pareto frontier using the highest BW/cap memory which satisfies the required capacity. 
Discrete jumps visible in the memory cost curve correspond to transitions between \mem{} tiers. 
While high BW/Cap memories are more expensive per GB, they eliminate capacity over-provision. 
As a result, total system cost is reduced.

Compared to using a fixed HBM3e memory, the \mem{} system reduces total cost by up to 12.4×. 
At scale, its memory-to-compute cost ratio matches that of an 8×H100 DGX~\cite{noauthor_groq_2024}, demonstrating that \mem{} enable efficient bandwidth scaling while keeping costs reasonable.

\begin{figure}
    \centering
    \includegraphics[width=\linewidth]{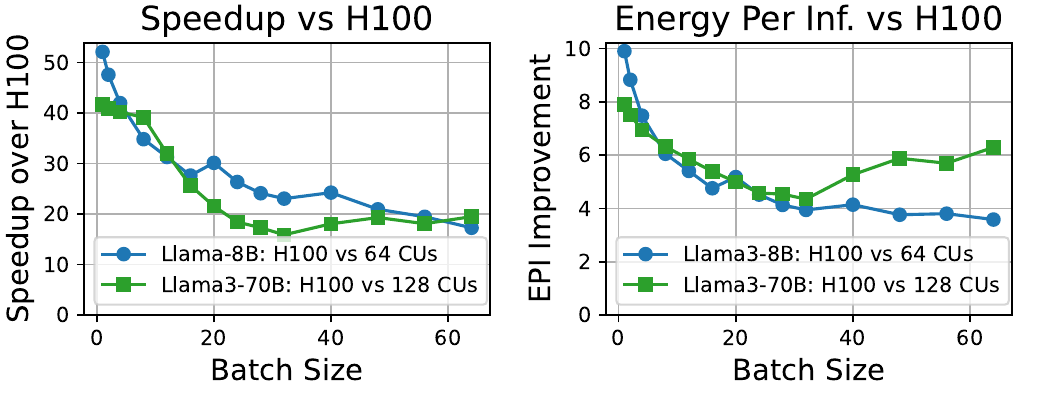}
    \caption{Speedup and energy-per-inference of an RPU versus and H100, sweeping batch size for Llama models with 8k prefill 2k decode.}
    \label{fig:on-chip-buffer}
    \label{fig:batch-scaling}
\end{figure}

\textbf{\textit{Energy per Inference versus Batch Size: }}
Figure~\ref{fig:batch-scaling} shows speedup and energy-per-inference of the RPU over an H100 across batch sizes for Llama3-8B and Llama3-70B.
Larger batch sizes improve the GPU’s compute efficiency.
However, concurrent queries inflate the KV\$ cache latency, introducing phase imbalance that the GPU cannot hide. 
For the RPU, small 4k sequences limit the benefit of decoupled pipelines because weight computation dominates, leaving less room to overlap KV\$ prefetching. 
As a result, performance gains plateau at $\sim$15-20× over the H100, though the RPU still maintains higher throughput and better energy efficiency.

At small batch sizes, the RPU shines, delivering over 40-50× speedup and 8-10x energy-per-inference, driven by its higher memory bandwidth and ability to efficiently execute small kernels with minimal synchronization overhead.
In contrast, H100 performance performs poorly in this regime, as it is significantly bandwidth-bound and suffers from kernel launch and scheduling overheads.

\section{Decomposed Contributions} \label{sec:decomposing-rpu-contributions}


\textit{Contribution 1 -- \mem{} Memory: }
Compared to an RPU system using HBM3e, \mem{} offers up to 2.2× lower energy per inference and 12.4× lower system cost, primarily by eliminating excess capacity and reducing internal wire lengths. 
These savings allow us to scale the number of compute units at ISO-TDP, leading to a 2.1× latency improvement.

\textit{Contribution 2 -- RPU Power and Area Provisioning:}
By rebalancing the compute-to-bandwidth ratio relative to an RPU provisioned like an H100 ($\sim$200 Ops/Byte), the RPU saves 3.3× die cost and 2.6× TDP utilization, leading to a 2.2× latency improvement when scaling out at ISO-TDP. 

\textit{Contribution 3 -- Microarchitectural Decoupling: }
Fine-grained network sharding eliminates global synchronization, avoiding up to a 2.0× latency penalty from collective stalls. 
Memory-compute decoupling enables deep prefetching, preventing a 1.2× slowdown from serialized kernel execution. 
In batch size 32 workloads, decoupled execution allows the RPU to straddle the roofline across memory-bound (SDPA, MoE) and compute-bound (Linear) kernels, improving latency by up to 1.6×.
These changes also improve energy efficiency: 1.4× over a monolithic NUMA-style baseline via shorter data paths, and 1.7× at the SRAM interface through on-the-fly stream dequantization. 
Together, these energy efficiency gains enable the system to scale to 2.4× more bandwidth at ISO-TDP.

\textit{Cumulative Performance:}
\mem{}, aligned provisioning, and decoupled pipelines enable 20-40× higher effective memory bandwidth at ISO-TDP, consistent with simulation results.

\textit{RPU Application Domain: }
Human-computer interaction literature identifies an interaction-latency threshold on the order of ten seconds, beyond which working memory decays and users are likely to context-switch, incurring re-orientation overheads~\cite{nielsen_slow_2025, maslych_mitigating_2025, gnewuch_opposing_2022}. 
Accordingly, reasoning systems must minimize end-to-end latency to preserve turn-taking and cognitive continuity, rather than maximizing throughput.
This captures the motivation behind the RPU: we want advanced intelligence at our fingertips.
The RPU targets these reasoning-intensive, interactive workloads requiring end-to-end responses such as multi-step planning, problem solving, iterative coding, and writing assistance, which currently take tens of seconds to minutes on today's systems~\cite{semianalysis_inferencemax_nodate}.
By exploiting the latency benefits of low-batch inference, it delivers state-of-the-art responsiveness and per-query performance.




\section{Related Work} \label{sec:related-work}
\textbf{\textit{DRAM-Centric General-Purpose Accelerators: }}
Systems such as NVIDIA H100~\cite{choquette_nvidia_2023}, AMD MI300x~\cite{smith_amd_2024}, SambaNova~\cite{prabhakar_sambanova_2024}, and TPU~\cite{jouppi_tpu_2023} use high-capacity HBM, large shared caches, and dense compute to support both training and inference. 
These architectures typically feature a single NUMA domain with distributed controllers and centralized caches. 
While this enables flexible data access, it creates long memory paths and high energy per access, which is especially harmful for memory-bound decode.
In contrast, the RPU uses a fine-grained NUMA design where each core has its own \mem{} DRAM channel and local SRAM buffer, eliminating shared caches and reducing on-chip data movement. 
This decouples compute, memory, and network pipelines, sustaining high bandwidth utilization.


\textbf{\textit{SRAM-Centric Custom Accelerators: }}
Custom accelerators such as Groq~\cite{noauthor_groq_2024}, Cerebras WSE-3~\cite{wang_cerebras_2024}, and Graphcore IPU~\cite{knowles_graphcore_2021} rely on SRAM as main memory. 
However, the limited density of SRAM makes it impractical to store large models efficiently. 
For example, a 70B parameter model deployed on Groq requires hundreds of accelerator cards, while Cerebras spans four wafer-scale chips.

To utilize their full SRAM bandwidth, these systems shard each matrix across a large compute fabric. 
For example, Cerebras may distribute a single VMM across 900,000 cores, requiring vector broadcasts to traverse up to 1,000 core-to-core hops and reductions to span the entire wafer. 
With the model globally distributed, network communication, not compute or memory access, becomes the primary performance bottleneck due to SRAM’s low density.
In contrast, each RPU reasoning core is significantly more capable than the ultra-lightweight cores used in Groq or Cerebras, with higher FLOP throughput and wider data buses. 
As a result, more of the workload is processed locally per core, reducing reliance on multi-hop communication. 
Additionally, the RPU’s hierarchical ring network has a much smaller diameter than mesh or wafer-scale fabrics, further minimizing the number of hops required for vector broadcasts and reductions.

\textbf{\textit{Processing In Memory (PIM) Accelerators: }}
PIM architectures aim to reduce data movement by embedding compute capabilities within or near memory~\cite{heo_neupims_2024, lee_hardware_2021, park_attacc_2024, zhou_transpim_2022, choi_unleashing_2023}. 
Many PIM designs leverage DRAM or emerging memory technologies to perform simple operations, typically integer or bitwise logic, in situ. 
While effective for low-intensity workloads, PIM designs struggle when arithmetic intensity exceeds 1 Op/Byte, which is common during LLM inference.
PIM architectures are also poorly suited for floating-point operations or fine-grained programmability to support rotary embeddings, softmax, and normalization functions.

Furthermore, the rise of block-quantized formats (e.g., BFP, MXFP) poses a major challenge for PIM. 
These formats require dynamic exponent broadcasting, alignment, and decoding before arithmetic. 
These steps involve conditional logic and variable indexing, which are difficult to implement in DRAM-compatible circuitry.


\begin{figure}
    \centering
    \includegraphics[width=\linewidth]{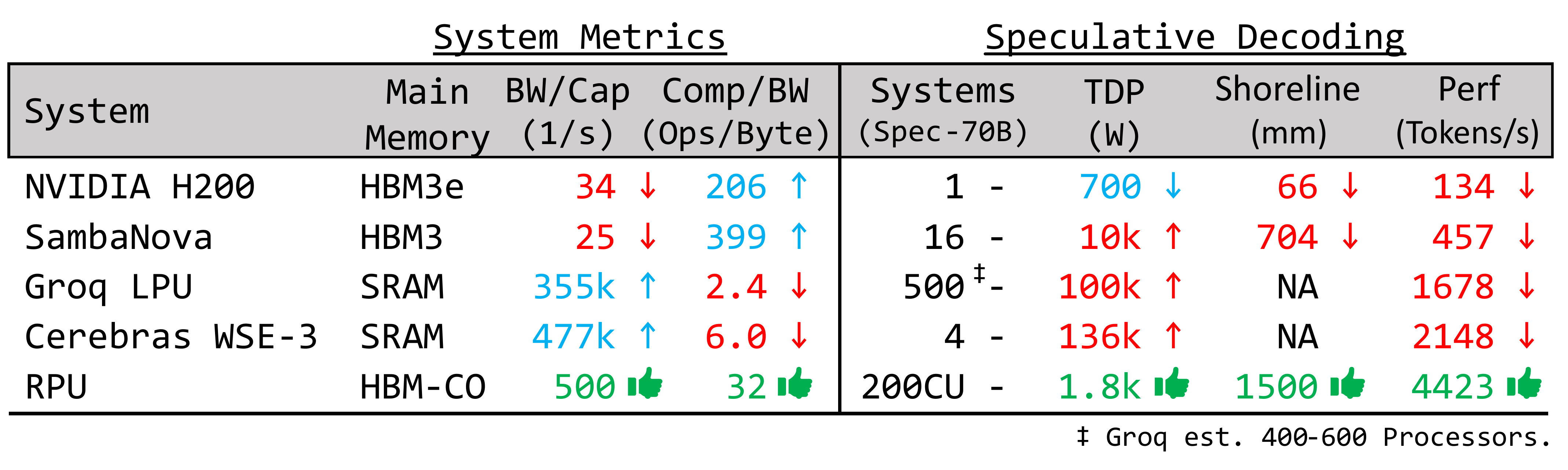}
    \vspace{-2em}
    \caption{
    A comparison of leading hardware platforms.
    Speculative decoding throughput for Llama3-70B based on published data~\cite{shah_boost_2024,  prabhakar_sambanova_2024, noauthor_groq_2024, wang_cerebras_2024}.
    }
    \label{fig:spec-decode-perf}
\end{figure}

\textbf{\textit{Comparison Under Speculative Decoding: }}
\label{spec-decode}
Speculative decoding is an increasingly common technique used in LLM inference to reduce token generation latency by leveraging a lightweight ``draft'' model to predict multiple tokens ahead. 
These predicted tokens are then validated by a larger ``target'' model; if the predictions are correct, several tokens can be committed in parallel. 
This approach may be challenging 
because it increases the arithmetic intensity of each query.

Industry accelerators often report performance under speculative decoding. 
We evaluate the RPU using a comparable speculative decode setup.
In our evaluation, we adopt an 8-token lookahead configuration in which a Llama3-8B draft model proposes tokens for a Llama3-70B target model. 
On average, 4.6 tokens are accepted per speculative window~\cite{mamou_accelerating_2024}, accelerating end-to-end inference by 1.8×. 
Figure~\ref{fig:spec-decode-perf} compares our speculative performance to publicly reported numbers from NVIDIA H200~\cite{shah_boost_2024}, SambaNova SN40L~\cite{prabhakar_sambanova_2024}, Groq~\cite{noauthor_groq_2024}, and Cerebras WSE-3~\cite{wang_cerebras_2024}. 
The RPU-200U configuration is lower latency than all evaluated systems.


\section{Conclusion}

Just as custom logic design led to configurable ASIC toolchains and foundry ecosystems, memory must follow a similar path.
Commodity DRAMs no longer aligns with the needs of modern inference systems.
Memory should be treated not as fixed infrastructure, but as a key dimension of system specialization.
With chiplet-based integration, customizable memory is not only desirable but also practical.
It allows system designers to co-optimize bandwidth, capacity, and energy by tailoring the design, packaging, and tuning of DRAM components for specific workloads.
The RPU system embraces this philosophy by co-designing its memory architecture with compute and interconnect, treating memory not as a constraint but as an opportunity for specialization.

\bibliographystyle{IEEEtranS}
\bibliography{references}

\end{document}